\documentclass[aps,prd,groupedaddress,showpacs,showkeys]{revtex4} 
\usepackage{latexsym,epsfig,amsmath,amssymb} 
\usepackage{amssymb} 
\usepackage{amsmath} 
\usepackage{amsfonts} 
\usepackage{epsfig} 
\usepackage{verbatim} 
\newcommand{\eq}{\begin{eqnarray}} 
\newcommand{\en}{\end{eqnarray}} 
 
\begin{document} 

\title{
Multiplicity distributions in $pp/p\bar{p}$ and $e^{+}e^{-}$ collisions\\
with parton recombination}

\author{ I.~Zborovsk\'{y}\footnote{Electronic address: zborovsky@ujf.cas.cz}}

\address{ Nuclear Physics Institute,\\
Academy of Sciences of the Czech Republic, \\
\v {R}e\v {z}, Czech Republic}

\begin{abstract}
A new approach to phenomenological description of the charged
particle multiplicity distributions in
$pp/p\bar{p}$ and $e^+e^-$ collisions is presented.
The observed features of the data  
are interpreted on the basis of stochastic-physical ideas of multiple production. 
Besides the processes of parton immigration and absorption, 
two and three patron incremental and decremental 
recombinations are considered. 
The complex behaviour of the multiplicity distributions at different energies 
is described by four parametric generalized hypergeometric 
distribution (GHD). 
Application of the proposed GHD to data  
measured by the CMS, ALICE, and ATLAS
Collaborations suggests that soft multiparton recombination processes 
can manifest itself significantly in the structure of multiplicity distribution 
in $pp$ interactions at very high energies.     

\end{abstract}

\pacs{13.85.Hd, 13.66.Bc}

\keywords{multiplicity distribution, high energy, parton recombination,
cascade processes }

\maketitle

{\section{Introduction}}

The multiple production of particles in
high energy collisions received a lot of attention
during many years.
A renewal interest in this topic originated recently with the beginning of operation
of the Large Hadron Collider (LHC) at CERN.
Among first results obtained at the LHC were multiplicity measurements
in proton-proton collisions. The CMS, ALICE,  and
ATLAS Collaborations provided valuable and precise data on multiplicity distributions (MD)
of the charged hadrons in the new super high energy domain.
The study of multiparticle production can give important information about
parton processes and transitions of the colored partons to the
colorless hadrons.
The complex phenomena that influence MD are however very hard to describe in detail,
because we cannot formulate the multiparton process in terms of
its QCD theory. Most of the particles which contribute to the
multiplicity buildup are soft particles and a perturbative
approximation is inapplicable here. Second, hadronization should be somehow
taken into account which is far of being understood properly.
The natural and probably most economical approach is to look for empirical
relations and regularities.

The study of particle production as a function of multiplicity
has revealed a very popular Koba-Nielsen-Olesen (KNO) scaling \cite{KNO}.
In $pp/p\bar{p}$ collisions, the scaling 
in the full phase-space holds up to the highest energy of the CERN 
Intersecting Storage Rings (ISR) 
but it is clearly violated \cite{UA5a}
in the energy region of the CERN Super Proton Synchrotron (SPS) collider and beyond.
The strong violation of KNO scaling 
was observed also at the energy $\sqrt{s}=7$~TeV 
in the limited pseudorapidity intervals $(|\eta|<2.4)$ though for 
small pseudorapidity
windows $(|\eta|<0.5)$ the scaling is approximately valid \cite{CMS}.
A related phenomenon is the so-called negative
binomial regularity,  which is the occurrence of the negative
binomial distribution (NBD) in different interactions
over a wide range of the collision energies. 
The UA5 Collaboration showed  
that MD in the non-single-diffractive (NSD) $pp/p\bar{p}$ collisions
can be described by NBD up to the energy $\sqrt{s}=546$~GeV  
both in the full phase-space \cite{UA5b}
and in symmetric pseudorapidity windows \cite{UA5c}.
Analysis of data on MD measured by the ALICE Collaboration 
indicates \cite{ALICE1,Mizoguchi}
that NBD describes the data in the small pseudorapidity window $|\eta|<0.5$ 
up to the energy $\sqrt{s}=2360$~GeV.
Besides $pp/p\bar{p}$ collisions, NBD  has been applied to 
various systems including $e^+e^-$ annihilations. 
Much effort has been made to explain the negative binomial form of MD observed   
in many situations, however, its physical origin has not been fully 
understood \cite{Grosse}.

The shape of MD of particles produced in hadron-hadron collisions
at high energies is quite different.
The full phase-space data on charged particle multiplicities
obtained from the NSD events in $p\bar{p}$
collisions at $\sqrt{s}= 900$~GeV \cite{UA5d}
indicated that besides KNO, the
negative binomial regularity is violated as well.
The measurements of MD at the energy  $\sqrt{s}= 1800$~GeV
by the E735 Collaboration \cite{E735} at the Fermilab Tevatron 
showed even stronger deviation of data from NBD. 
Though with large mutual discrepancies at high multiplicities,
both data demonstrate a narrow
peak at the maximum and some structure around $n\sim~2<n>$. Despite
its correct qualitative behaviour, the single NBD is not sufficient to
describe the experimental data.
A systematic study of the complex form of MD was performed in the framework of
two component model \cite{GiovUgo1} and its three component modification \cite{GiovUgo2}.
There exist also other approaches to explanation of the observed features of particle 
production at high energies (for a review see Ref. \cite{Grosse}).  
New interest in this field is motivated by the recent results of the multiplicity measurements 
in $pp$ collisions at the LHC. The data obtained by the CMS \cite{CMS}, 
ALICE \cite{ALICE2}, and ATLAS \cite{ATLAS}
Collaborations show similar structure in MD of charged particles produced 
in the limited windows in pseudorapidity.   
The measurements allow to study evolution of the distinct peak 
at maximum and the broad shoulder at large $n$  
both with the collision energy and pseudorapidity.

In this paper we propose an alternative phenomenological approach to 
the description of MD in $pp/p\bar{p}$ and $e^{+}e^{-}$ collisions.
Using the same concept for hadron and lepton interactions,
we try to account for the structure in data
which emerges in the $pp/p\bar{p}$ interactions at high energies.
The proposed four parametric representation of MD is motivated by a scenario of
the parton cascading.
The considerations behind are based on a premise that dynamics
of the particle production as manifested on the level of
multiplicities can be described in terms of a stochastic
cascading with specific types of the underlying processes.
The construction includes the processes of parton immigration and absorption
together with two and three parton incremental and decremental recombination.
The recombination processes are assumed in the final stage 
of the parton evolution during the color neutralization. 
The transition to the colorless hadrons is considered in a stationary regime  
at breakdown of the confinement and onset of hadronization.

\vspace*{0.8cm}
{\section{Recurrence relation between multiplicities n and n+1}}

The number of particles created in high energy collisions varies from event to event.
The distributions of probabilities $P_n$ of occurrence of the multiplicity $n$
provide sensitive means to probe dynamics of the interaction.
Besides general characteristics
of particle production the distributions contain
information about multiparticle correlations in an integrated form.
The correlations of the final particles 
reflect features of hadronization mechanism and properties 
of the QCD parton evolution just before hadronization.  
The data on MD of charged particles from the high energy 
$pp/p\bar{p}$ and $e^+e^-$ interactions 
allow us to study the processes underlying 
the multiple production and its correlation structure     
within a phenomenological framework of parton cascading.
The extraction of information on these processes   
requires the elementary charge conservation to be taken into account.  
In our approach we investigate and exploit MD of charged particle pairs 
and characterize it by a recurrence relation between $P_n$ and $P_{n+1}$. 
This assumes connection between the collisions
of multiplicities $n+1$ with $n+1$ collisions of multiplicity $n$ which can be written
in the form
\begin{equation}
\frac{(n+1)P_{n+1}}{P_{n}} = g(n).
\label{eq:r1}
\end{equation}
The class of MD defined by expression of this type has been considered with a view
to stimulated emission and cascading in Ref. \cite{GiovanniniVanHove}.
The independent emission of particles represented by Poisson distribution
is characterized by $g(n)=c$. The constancy of $g(n)$ means that creation of an
additional particle is independent of number of other particles.
The stimulated emission of identical bosons obeying Bose-Einstein statistics
follows geometric distribution for which one has $g(n)=c(n+1)$.
This means that emission probability of a boson is enhanced by a factor $n+1$ when $n$ bosons
are already present in the system.
Both examples are special cases of NBD given by the formula
\begin{equation}
P_n=\frac{(n+k-1)!}{n!(k-1)!}q^{n}(1-q)^k.
\label{eq:r2}
\end{equation}
The distribution depends on two parameters $q=<n>/(<n>+k)$ and $k$ for which
the recurrence relation (\ref{eq:r1}) is given by the linear dependence
\begin{equation}
g(n) = (k+n)q.
\label{eq:r3}
\end{equation}
For fixed $<n>$, the Poisson distribution is recovered with $q,k^{-1}\rightarrow 0$.
The geometric distribution corresponds to $k=1$.
Chew et al. proposed \cite{Chew} a generalization of NBD  
which gives good description of MD
of charged particles in $e^{+}e^{-}$ annihilation \cite{Dewanto}.
The generalized multiplicity distribution (GMD) has the form
\begin{equation}
P_n=\frac{(n+k-1)!}{(n-k')!(k-1)!}q^{n-k'}(1-q)^{k+k'}.
\label{eq:r4}
\end{equation}
The distribution is function of three parameters $q=(<n>-k')/(<n>+k)$, $k$, and $k'$.
The expression for $g(n)$ in this case reads
\begin{equation}
g(n) = \frac{(n+1)(k+n)}{(n+1-k')}q.
\label{eq:r5}
\end{equation}
The formula for GMD reduces to the expression for NBD, when $k'=0$.
For $k=0$, GMD becomes Furry-Yule distribution (FYD) proposed by Hwa and Lam \cite{HwaLam}.
The mentioned distributions are fully determined by the functional form of $g(n)$
together with the general normalization condition to unity. 
The normalization is guaranteed for $n>k'-1$ by
the linear dependence of $g(n)$ at large multiplicities 
with $q< 1$.
In the parton cascade picture of multiple production 
the parameters of these distributions can be related with the rate constants 
of branching processes in the evolution equations for probabilities.

\vskip 0.8cm
\subsection{Cascade processes with parton recombination}

In a high energy collision with the creation of a multiparticle
state the dynamics of the system can be simulated as a parton
cascading associated with quark and gluon interactions during the
interaction time. The corresponding parton cascade
processes inspired by the elementary perturbative QCD are of three
types: a/ parton-parton collisions, b/ branching processes such as
quark bremsstrahlung, gluon self interaction or gluon splitting, c/
fusion processes e.g. gluon annihilation on a quark.
Practically all QCD-based models of many-body production involve
some form of cascading. The stochastico-physical
picture of the multiplicity evolution can be described by
Kolmogorov-Chapman differential-difference (DD) rate equations \cite{Feller}
for the probability $P_{n}(t)$. The continuous parameter
$t$ is usually interpreted as an ordering parameter,
QCD evolution parameter, or time. 
The rate equations were applied to
hadron physics many years ago in Ref. \cite{Giovannini1}. 
Soon afterwards, the properties of MD were studied by
solving DD equations in terms of parton cascading 
by number of people \cite{Biyajima}.
In particular, all possible QCD vertices were taken into account in Ref. \cite{Durand}.
Stationary regime in birth and death processes with recombination (confluence) 
of two ($2\rightarrow 1$) and three ($3\rightarrow 1$) gluons has been investigated 
in Ref. \cite{Batunin}.

In this paper we point out and exploit somewhat different strategy. We try to establish 
minimal number of such processes in the cascade picture of particle 
production which could account for both the general character of MD and 
its complex structure observed at high energies.
A successful description of experimental data
requires essential features of the cascade evolution to be introduced
into the phenomenology.
We attempt to estimate those features  
which mostly influence the building structure of MD at high energies.
This involves processes of parton recombinations at the end
of the cascade just near breakdown of the confinement.
The last phase of the parton evolution is 
considered to have large impact on the form of MD of the produced hadrons.
We suppose that in the final stage 
there exists some kind of a stationary regime which occurs at the transition between
parton and hadron degrees of freedom.
In this regime, soft partons intensively exchange their momenta to reach an 
"momenta uniformity" before their conversion into observable particles \cite{Batunin}.

In order to introduce features of parton recombination into a multiplicity 
description, we consider a stationary regime of DD equations of Kolmogorov-Chapman type.
Hereafter we refer to the partons as objects
which cascade in "time" and regard the terms particle and parton 
as interchangeable.  The binary parton-parton
scatterings do not change the number of particles during the system evolution 
but can influence neutralization or screening of the color flow. 
They can act as supporting processes which initiate
branching or fusion of the nearby partons.
Such connection represents multiparton interactions contributing to 
a change of the particle number by one.
The simplest interactions of this type are recombinations of two or three particles.
Here we consider the process of two parton incremental recombination ($2\rightarrow 3$) 
together with three parton incremental ($3\rightarrow 4$) 
and three parton decremental ($3\rightarrow 2$)
recombination processes.
The confluence of two and three partons, $2\rightarrow 1$ and $3\rightarrow 1$, 
is supposed to be negligible  
with respect to other recombination processes at the end of the cascade.
Together with the particle production that depends 
on particles already produced, we consider independent immigration ($0\rightarrow 1$) 
and absorption ($1\rightarrow 0$) as another source of partons 
influenced by bulk properties of the system created in the collision. 
The accumulated energy is transformed
into partons in this manner, or on the contrary a parton can be
melted and absorbed by the expanding system again.
In the case of independence of each cascade process, the corresponding DD evolution equations
for the probability $P_{n}(t)$ read
\begin{eqnarray}
\dot P_n = &+&\alpha_{0} P_{n-1}
                      - \alpha_{0} P_n\nonumber\\
                     &+&\beta_{0}(n\!+\!1) P_{n+1}
                      - \beta_{0}n P_n\nonumber\\
                     &+&\alpha_{2}(n\!-\!1)(n\!-\!2)P_{n-1}
\vspace{.85in}
                      - \alpha_{2}n(n\!-\!1)P_n\nonumber\\
                     &+&\alpha_{3}(n\!-\!1)(n\!-\!2)(n\!-\!3)P_{n-1}
\vspace{.85in}
                      - \alpha_{3}n(n\!-\!1)(n\!-\!2)P_n\nonumber\\
                     &+&\beta_{2}(n\!+\!1)n(n\!-\!1)P_{n+1}
\vspace{.85in}
                      - \beta_{2}n(n\!-\!1)(n\!-\!2)P_n .
\label{eq:r6}
\end{eqnarray}
The coefficients $\alpha_0$, $\alpha_2$, $\alpha_3$ are the corresponding
rates for the processes $0\rightarrow 1$, $2\rightarrow 3$,
$3\rightarrow 4$, respectively. Similar parameters for degradation 
of the particle number
in the processes $1\rightarrow 0$
and $3\rightarrow 2$ are denoted as $\beta_0$ and
$\beta_2$. Using the definition of the generating function
\begin{equation}
Q(w) = \sum_{n=0}^{\infty} P_n w^{n},
\label{eq:r7}
\end{equation}
we rewrite the system of DD equations (\ref{eq:r6}) in terms of $Q(w)$.
The corresponding stationary  solution $(\dot P_n=0)$ satisfies the third order
differential equation
\begin{equation}
\alpha_0Q(w) -  \beta_0\frac{dQ(w)}{dw} +
\alpha_2w^{2}\frac{d^{2}Q(w)}{dw^{2}} +
(\alpha_3w^3-\beta_2w^{2})\frac{d^{3}Q(w)}{dw^{3}}  = 0.
\label{eq:r8}
\end{equation}
A regular solution of this equation is proportional to the
generalized hypergeometric function
\begin{equation}
Q(w) = \ _{3}F_{2}\left(a_{1},a_{2},a_{3};b_{1},b_{2}.
       \frac{\alpha_{3}}{\beta_{2}}w\right)
\label{eq:r9}
\end{equation}
The complex constants $a_i$ and $b_j$ are given in terms of the real
parameters $\alpha_0$, $\alpha_2$, $\alpha_3$, $\beta_0$, and $\beta_2$ (see Appendix A).
The ratio (\ref{eq:r1}) is given as follows
\begin{equation}
\frac{(n+1)P_{n+1}}{P_{n}} =
\frac{\alpha_{0}+\alpha_{2}n(n-1)+\alpha_{3}n(n-1)(n-2)}{\beta_{0}+\beta_{2}n(n-1)}.
\label{eq:r10}
\end{equation}
This recurrence relation together with the normalization condition for $P_n$
fully determines the formula for MD which we
refer to as the generalized hypergeometric distribution (GHD).
The distribution depends on four
parameters $\alpha_0/\beta_2$, $\alpha_2/\beta_2$, $\alpha_3/\beta_2$, and
$\beta_0/\beta_2$ which are ratios of the rate constants for the corresponding
cascade processes.

In the next section we exploit GHD to describe MD of the charged particles
produced in $pp/p\bar{p}$ and $e^{+}e^{-}$ interactions.
The formula (\ref{eq:r10}) has various limits applicable for 
both reactions at different energies and phase space regions.
If $\alpha_0\rightarrow 0$ and $\beta_0\rightarrow 0$ simultaneously, GHD reduces to
NBD. In that case $g(n)$ is given by (\ref{eq:r3}) with the parameters
$q=\alpha_3/\beta_2$ and $k=\alpha_2/\alpha_3-2$. On the contrary, if
$\alpha_0$ and $\beta_0$ are both large enough relative to other parameters, a Poisson-like peak
appears in MD at small $n$ for which $g(n)\simeq \alpha_0/\beta_0$.
There exists a region
where GHD can be narrower than Poisson distribution. This happens
at low energies where the rate constants $\alpha_2$ and $\alpha_3$ are small or even vanish.

\section{Analysis of Data}

The statistics of the charged particle MD is governed by the
requirements of conservation laws in each collision.
Due to the global charge conservation, there are only even multiplicities when
dealing with experimental data in the full phase-space.
On the other hand the charge conservation cannot be satisfied by the stochastic processes which are independent of the charge.
In particular, GHD, as defined for all
non-negative integer values of $n$, cannot directly represent the full
phase-space data on multiplicities which are always even.
There are two main conceptions how to tackle the problem.
According to the often used procedure \cite{UA5b},  
the probabilities for only even integers $n$ are taken from a theoretical distribution,
then are renormalized and compared to the measured values of $P_n$.
The second approach assumes to deal with MD of particle pairs
and compare it to a theoretical distribution for all values of $n$.  
Both methods give different values of the parameters entering the distribution 
and differ also in capability of description of experimental data. 
Here we use the second method and apply GHD and NBD (for comparison)  
to the distribution of particle pairs.

We have analysed experimental data on MD of charged particles produced 
in the non-single-diffractive (NSD) $pp/p\bar{p}$ collisions in the full phase-space.  
The analysis was performed with the high energy data measured by the E735 Collaboration \cite{E735} 
at $\sqrt{s}=1800$, 1000, 546, and 300~GeV, by the UA5 Collaboration \cite{UA5} at 
$\sqrt{s}=900$ and 200~GeV, and by the ABCDHW Collaboration \cite{ABCDHW} at 
$\sqrt{s}=63$, 53, 45, and 30~GeV. The study includes also the FNAL fixed target 
data \cite{F405,F300,F200,F100}
at $\sqrt{s}=27.6$, 23.8, 19.7, and 13.8~GeV corrected for diffraction cross sections
\cite{FNALdif}, the Serpukhov data at $\sqrt{s}=11.5$~GeV \cite{Serpukhov} with diffraction 
corrections \cite{SerpukhovDIF}
and the low energy CERN data \cite{Blobel} measured by the BHM Collaboration 
at $\sqrt{s}=6.8$ and 4.9~GeV.
The results of the analysis with the CERN-MINUIT program
are shown in Table I and in Figs. 1-2. 
As can be seen from Table I, the description of MD of particle pairs by NBD is 
in most cases unsatisfactory including the data from ISR and lower energies.  
Though MD of the charged particles can be approximated by NBD 
up to the energy $\sqrt{s}=200$~GeV pretty well,  
NBD is not able to account fully for the narrower distribution of the particle pairs. 
At the energies lower than $\sqrt{s}\sim 20$~GeV, the distribution becomes 
even narrower than Poisson distribution. This corresponds to negative values 
of the parameters $k$ and $q$, meaning that NBD transforms to a binomial one.  
%
%
\begin{table}[htb!] 
\caption{The results of analysis of MD of the charged particle pairs in the NSD $pp/p\bar{p}$ 
collisions in the full phase-space.} 
\label{tab:pp fps} 
\begin{center} 
\begin{ruledtabular} 
\begin{tabular}{ccccccccccc} 
$\sqrt{s}$ & {} & NBD & {} & {} & {} &  GHD
\\ 
\cline{3-5}\cline{7-11}
{(GeV)} & {} &  k & q & $\chi^2/\mbox{NDF}$ & {} & 
$\alpha_0/\beta_2$ & $\alpha_2/\beta_2$ & $\alpha_3/\beta_2$ & $\beta_0/\beta_2$ &
$\chi^2/\mbox{NDF}$ 
\\ 
\hline 
1800  & E735  & $3.36 \pm 0.04$ & $0.877 \pm 0.001$  & 670/126      & {} &
$(2.7 \pm 0.4)10^3$ & $17 \pm 1$ & $0.76\pm 0.01$ & $341\pm 41$ & 30.1/124
\\ 
1000  & E735   & $3.4 \pm 0.1$ & $0.854 \pm 0.003$  & 106/75       & {} &
$(5.0 \pm 2.1)10^3$ & $27 \pm 8$ & $0.57\pm 0.09$ & $600\pm 234$ & 54.5/73
\\ 
{ 900} & UA5  & $4.0\pm 0.1$ & $0.815\pm 0.005$  & 84.4/52       & {} &
$(1.2 \pm 0.9)10^3$ & $64 \pm 2$   & {-}           & $1545\pm 94$ & 11.0/51
\\ 
{ 546} & E735  & $5.0\pm 0.1$  & $0.776 \pm 0.002$  & 165/78      & {} &
$(2.5 \pm 1.2)10^3$ & $19 \pm 6$ & $0.61\pm 0.08$ & $339\pm 155$ & 29.8/76
\\ 
{ 300} & E735  & $5.4 \pm 0.2$ & $0.776 \pm 0.005$  & 52/58       & {} &
$(1.1\pm 1.0)10^3$ & $16 \pm 7$ & $0.49\pm 0.14$ & $193\pm 154$ & 14.3/56
\\ 
{ 200} & UA5  & $5.8 \pm 0.3$ & $0.64\pm 0.01$  & 13.3/29      & {} &
$(3.6\pm 0.7)10^2$ & $33  \pm 3$   & {-}           & $573\pm 96$ & 3.2/28
\\ 
{  63} & ISR & $21\pm 2$ & $0.25 \pm 0.02$  & 30.1/18         & {} &
$1.7 \pm 4.3$ & $5.3\pm 0.2$ & $0.29\pm 0.03$ & $0.2\pm 0.5$   & 19.8/16
\\ 
{  53} & ISR   & $24\pm 3$ & $0.21 \pm 0.02$  & 30/19       & {} &
$16\pm 15$     & $4.6\pm 2.3$ & $0.34\pm 0.04$ & $2.2\pm 2.3$ & 4.6/17
\\ 
{  45} & ISR  & $59\pm 20$ & $0.09 \pm 0.03$  & 44.9/17       & {} &
$15\pm 9$     & $4.1\pm 0.4$ & $0.36\pm 0.06$ & $1.5\pm 1.1$ & 4.5/15
\\ 
{  30} & ISR & $49\pm 21$ & $0.10 \pm 0.04$  & 22.5/15         & {} &
$0.5 \pm 2.0$ & $4.5\pm 0.3$ & $0.16\pm 0.05$ & $0.03 \pm 0.13$ & 4.5/13
\\ 
{  27.6}  & FNAL & $36\pm 17$ & $0.11 \pm 0.05$  & 22.1/14       & {} &
$5\pm 10$        & $3.8\pm 0.5$ & $0.22\pm 0.08$ & $0.6\pm 1.6$ & 15.5/12
\\ 
{  23.8} & FNAL & $(19 \pm 3)10^4$ & $(24 \pm 4)10^{-6}$  & 12.2/12       & {} &
$6\pm 10$       & $4.0\pm 0.4$ & $0.07\pm 0.07$ & $0.8\pm 1.8$ & 7.6/10
\\ 
{  19.7} & FNAL & $(-8\pm 5)10^3$ & $(-5 \pm 3)10^{-4}$  & 24/11       & {} &
$3.6\pm 3.6$    & $3.8\pm 0.4$ & $0.00\pm 0.07$ & $3.6\pm 3.6$ & 4.5/9
\\ 
{  13.8} & FNAL & $-13 \pm 1$ & $-0.35\pm 0.04$  & 46.5/8        & {} &
$4.6\pm 1.2$    & $2.8\pm 0.1$ & {-}           &  $0.5\pm 0.2$ & 8.0/7
\\ 
{  11.5} & Serpukhov & $-14 \pm 1$ & $-0.28\pm 0.03$  & 23.7/7       & {} &
$4.7\pm 2.0$         & $2.3\pm 0.1$ & {-}           & $0.8\pm 0.5$ & 8.8/6
\\ 
{   6.8} & CERN/BHM & $-5.0\pm 0.2$ & $-0.90\pm 0.10$  & 4.7/3 & {} &
$7.9\pm 5.0$        & $0.5\pm 0.4$ & {-}           & $1.3\pm 1.3$ & 0.2/2
\\ 
{   4.9} & CERN/BHM & $-4.1\pm 0.1$ & $-0.64\pm 0.03$  & 25.5/3 & {} &
$3.5\pm 0.3$        & {-}          & {-}           & $0.9\pm 0.2$ & 5.4/3
\end{tabular} 
\end{ruledtabular} 
\end{center} 
\end{table} 
By virtue of its four parameters, GHD describes the complex structure of the charged particle MD emerging in the high energy $pp/p\bar{p}$ collisions in the full phase-space sufficiently well. 
This is illustrated in Fig.1(a) on data \cite{E735} from the E735 Collaboration where a peak at 
low $n$ and a shoulder at large multiplicities is visible.  
The evolution of the observed structure with the energy $\sqrt{s}$ 
is shown in more detail in Fig.1(b).
Here the relative residues of MD with respect to the NBD parametrization are depicted.
The residues are mutually shifted  
by factors 2 for single energies.
The lines correspond to the description of data with GHD. 
As one can see from Fig.1(b), the shoulder broadens with 
the energy $\sqrt{s}$ and its maximum moves towards 
larger multiplicities. This corresponds to an increase of the parameter
$\alpha_3/\beta_2$ which is ratio of the rate constants for the 
processes $3\rightarrow 4$ and $3\rightarrow 2$. 
The measurements of MD by the UA5 collaboration at $\sqrt{s}=200$ and 900 GeV
in the full phase-space lay systematically below the E735 data \cite{E735}
for large $n$.   
The discrepancy has a consequence that the parameter $\alpha_3/\beta_2$
is negative for the UA5 data. Therefore we set it to null in our analysis  
in this particular case (see Table I).  

The description of data on MD by GHD in the NSD proton-proton collisions at the ISR 
energies is shown in Fig.2(a). The relative residues with respect to the NBD parametrizations
are depicted in Fig.2(b). The residues are mutually shifted  
by factor 2 for single energies.
One can see from Fig.2(b) that NBD does not represent accurate  
parametrization of MD of particle pairs even at the ISR energies.    
Especially for low $n$ the residues od data relative to NBD show remnants of
the peaky structure which is clearly visible in the TeV energy region. 
The solid lines represent the parametrization of data by GHD.   
The small experimental errors allow good determination of $\alpha_2/\beta_2$ and $\alpha_3/\beta_2$ in this region. 
The values of the parameters are non-zero
at the ISR energies but they are smaller than at the TeV energies.
Similarly, the parameters $\alpha_o/\beta_2$ and $\beta_0/\beta_2$ are non-zero 
though both are relatively small. This is once again a reformulation of the statement that 
NBD is not sufficient to describe MD of the particle pairs 
nevertheless at the ISR energies it provides much better approximation of data 
than in the TeV energy region.

The analysis of MD below $\sqrt{s}\sim 20$~GeV showed that the parameter 
$\alpha_3$ becomes negative. Therefore we set it equal to null in this region. 
It means that the process $3\rightarrow 4$ dies out as the first at low energies.
For still smaller $\sqrt{s}$ there is not enough energy even for 
the process $2\rightarrow 3$ and GHD becomes two parameter distribution.
In this region, MD of particle pairs is extremely narrow. 
The non-zero value of $\beta_2$ makes it narrower than the Piosson distribution. 
This means that the parton recombination process $3\rightarrow 2$ remains active till 
the very small energies 
though it brings along only the diminution of the particle number.     
%
%
\label{Figure:1}
\begin{figure}[ht!] 
\vskip 3.8cm
\begin{center}
\hspace*{-1.5cm}
\parbox{6cm}{\epsfxsize=5.cm\epsfysize=4.3cm\epsfbox[95 95 400 400]
{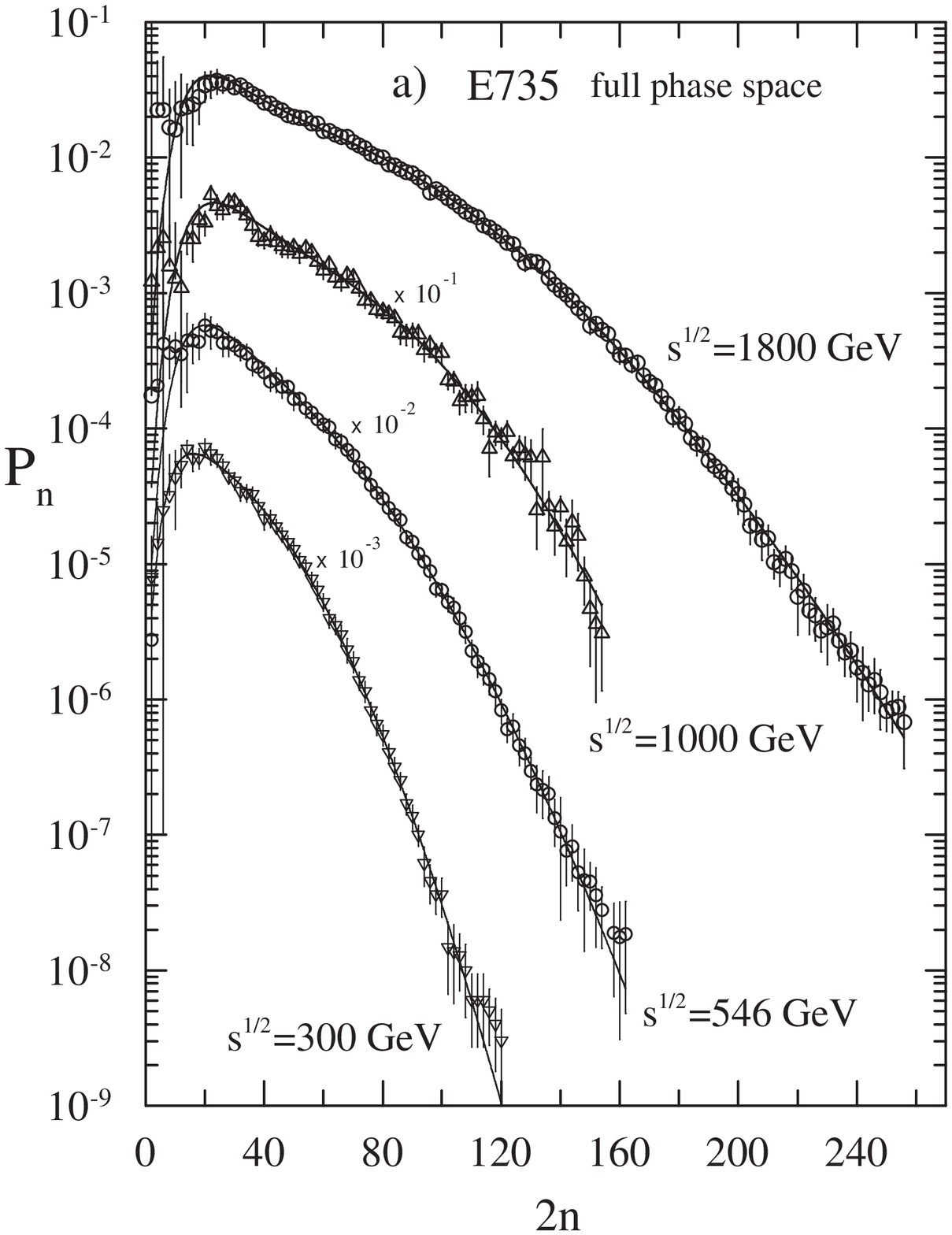}{}}
\hspace*{2.5cm}
\parbox{6cm}{\epsfxsize=5.cm\epsfysize=4.3cm\epsfbox[95 95 400 400]
{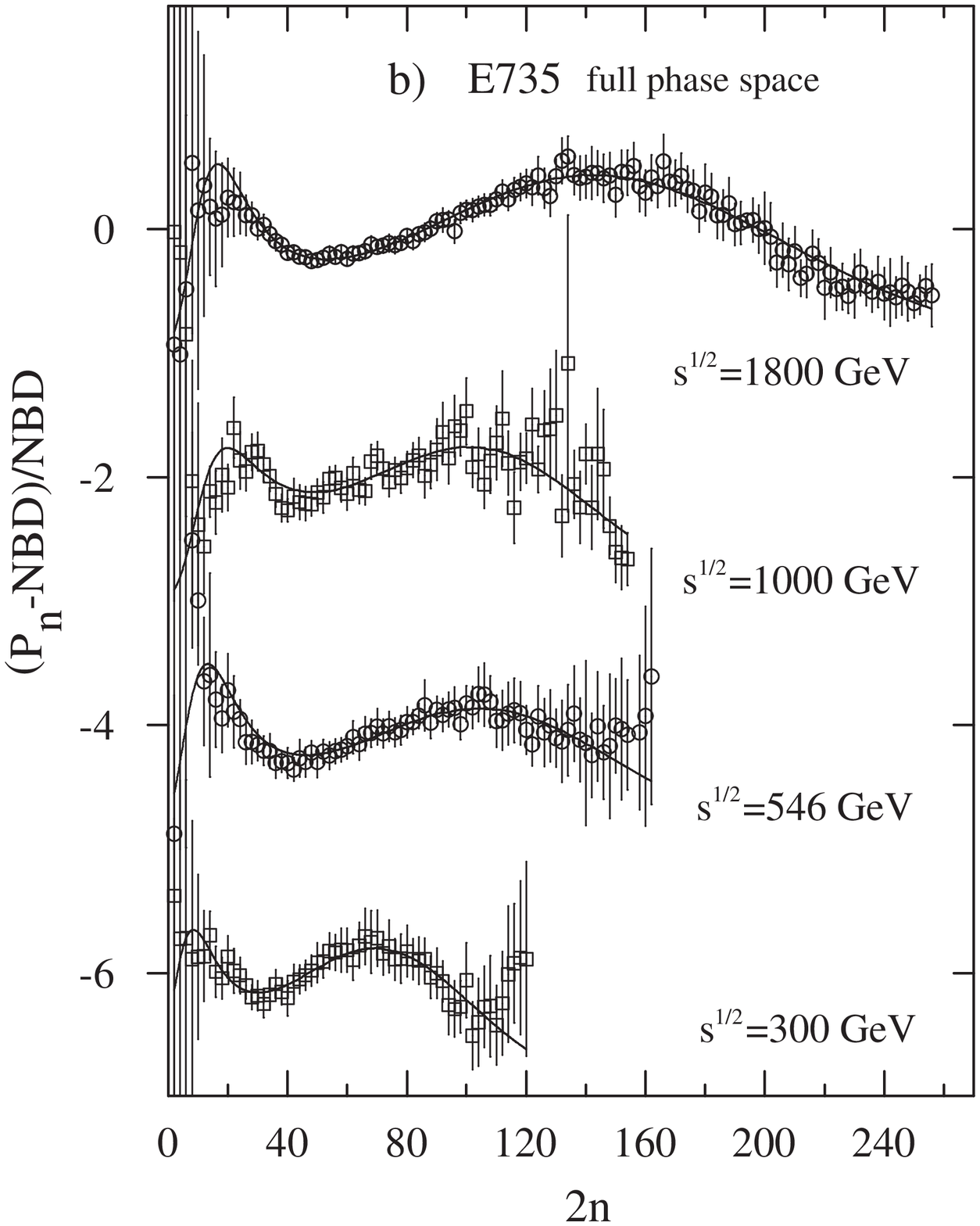}{}}
\vskip 1.5cm
\end{center}
\caption{ (a) MD of charged particles 
in the NSD $p\bar{p}$ collisions at energies $\sqrt s=300-1800$~GeV. The data \cite{E735} are 
multiplied by the indicated factors. 
(b) The relative residues of MD with respect to the NBD parametrization. 
The residues are mutually shifted by the factor of 2 at different energies. 
The lines represent description of the data by GHD. } 
\end{figure} 
%
%
\label{Figure:2}
\begin{figure}[h!] 
\vskip 3.8cm
\begin{center}
\hspace*{-1.5cm}
\parbox{6cm}{\epsfxsize=5.cm\epsfysize=4.3cm\epsfbox[95 95 400 400]
{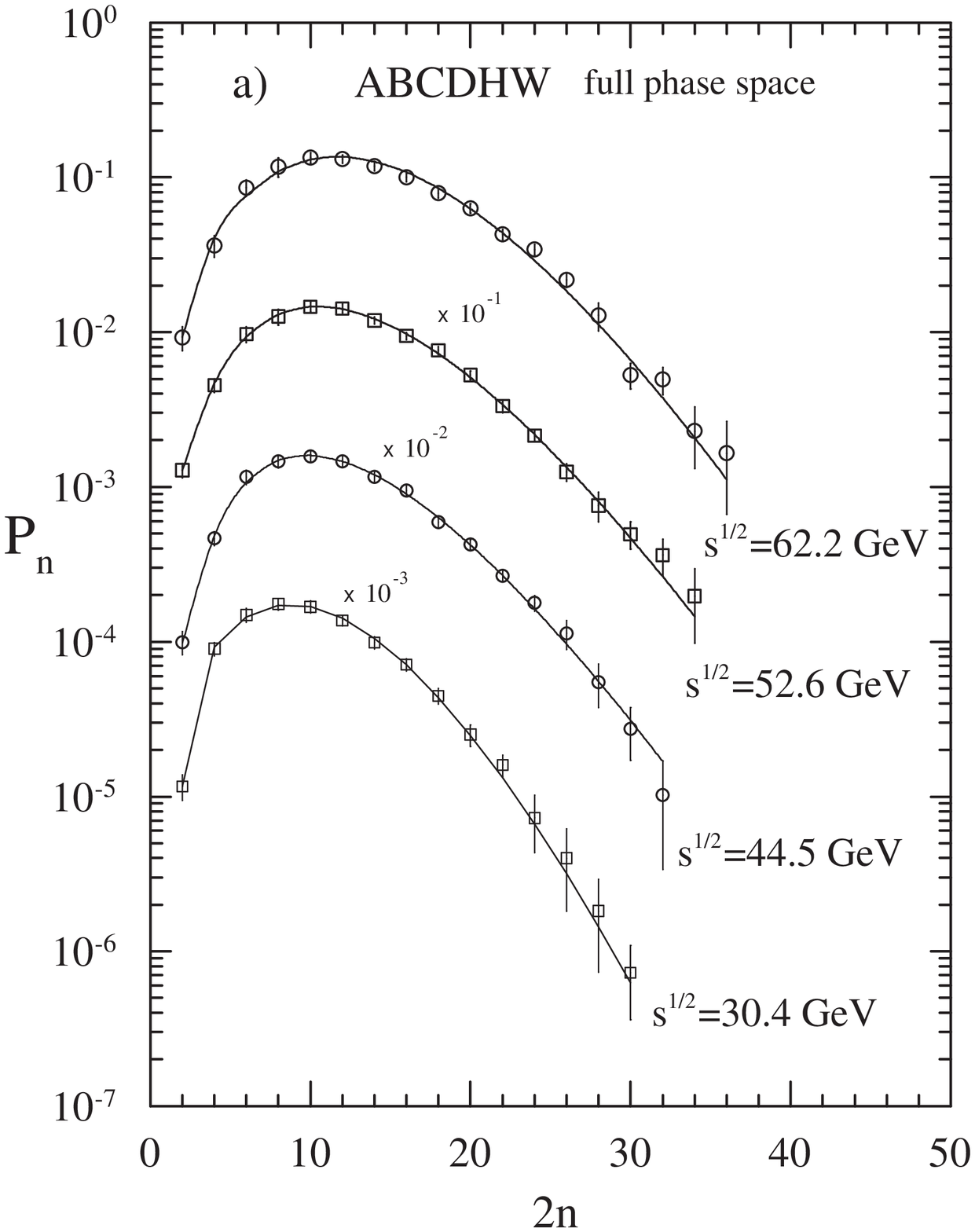}{}}
\hspace*{2.5cm}
\parbox{6cm}{\epsfxsize=5.cm\epsfysize=4.3cm\epsfbox[95 95 400 400]
{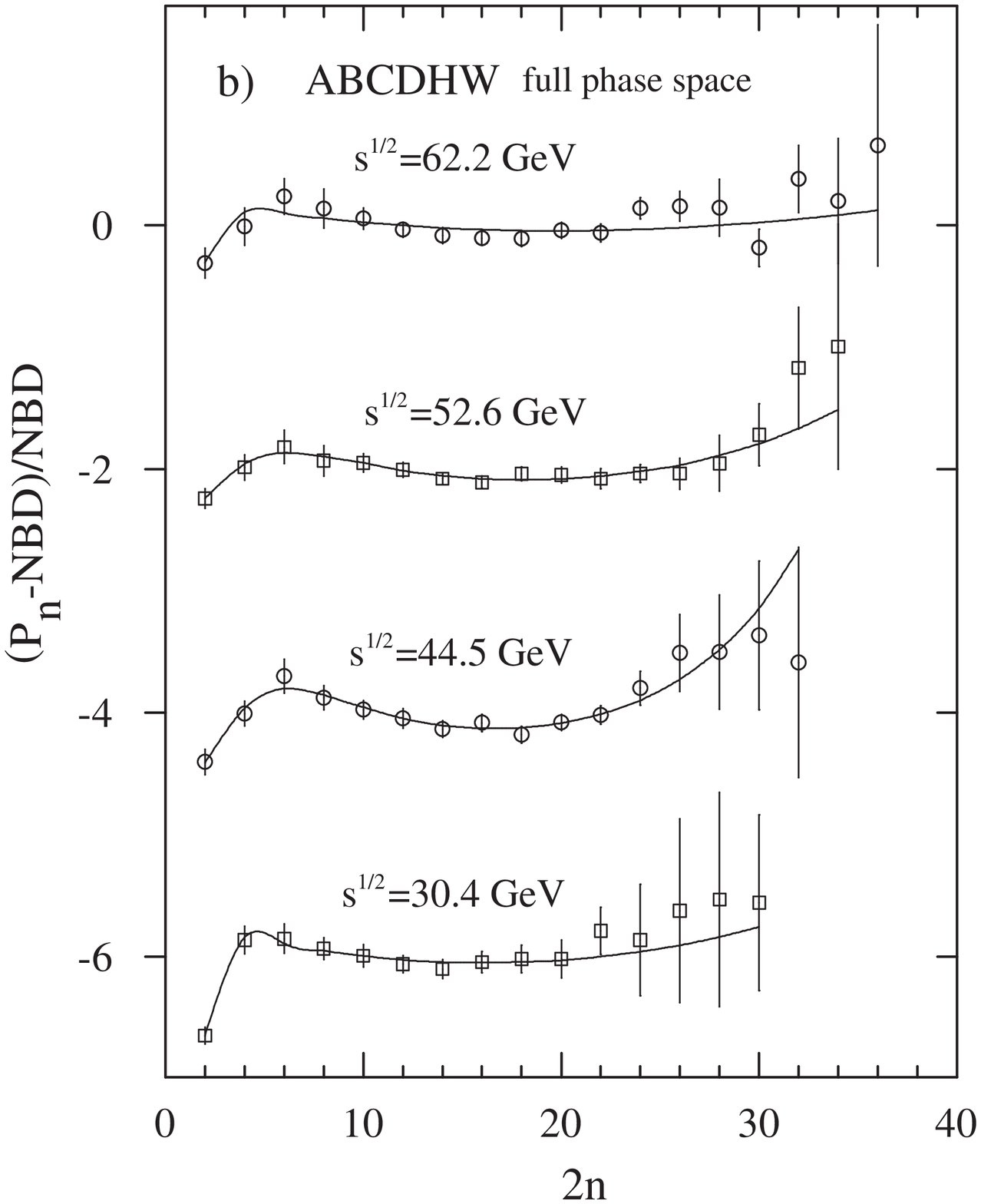}{}}
\end{center}
\vskip 1.2cm
\caption{
(a) MD of charged particles 
in the NSD $pp$ collisions at the ISR energies. 
The data \cite{ABCDHW} are 
multiplied by the indicated factors. 
(b) The relative residues of MD with respect to the NBD parametrization. 
The residues are mutually shifted by the factor of 2 at different energies. 
The lines represent description of the data by GHD. }
\end{figure} 

\subsection{Structure of multiplicity in limited phase-space regions}

Recently, the CMS Collaboration provided results of systematic measurements of 
charged particle multiplicities in $pp$ collisions at the LHC. The data \cite{CMS}
were accumulated in five pseudorapidity ranges from $|\eta|<0.5$ to 
$|\eta|<2.4$ at the collision energies $\sqrt{s}=900$, 2360, and 7000 GeV.
The measurements of MD in the restricted phase-space regions
can serve as a sensitive probe of the underlying dynamics 
in various phenomenological models.

For an adequate description one needs to
make additional assumptions when projecting the full phase-space data
onto the smaller pseudorapidity windows $|\eta|<\eta_c$. 
The odd-even effect of the charged particle 
distribution $P^{ch}_n$ is smeared out with the decreasing $\eta_{c}$ 
and the distribution fills all values of $n$.
We avoid the complication concerning the charged particle measurements 
in the limited phase-space regions and collect the data in the neighbouring 
even and odd bins as follows 
\begin{equation}
P_{n} = P^{ch}_{2n}+P^{ch}_{2n-1}, \ \ \   n=1,2...,
\label{eq:r11}
\end{equation}
so simulating the distribution of particle pairs.
It would be more correct to deal with MD of negative particles
which is essentially the distribution of the charged particle pairs. 
Nevertheless, application of GHD to the distribution (\ref{eq:r11}) 
allows us to establish main trends which characterize MD of negative particles.
Here we study the dependence of the parameters $\alpha_0/\beta_2$,
$\alpha_2/\beta_2$, $\alpha_3/\beta_2$, and $\beta_0/\beta_2$ 
on the size of the pseudorapidity span $|\eta|<\eta_c$.

In Fig. 3(a) we show MD of charged particles in the limited phase-space regions
measured by the CMS Collaboration in $pp$ collisions at $\sqrt{s}=7000$~GeV.
The CMS data for the five pseudorapidity ranges from $|\eta|<0.5$ to 
$|\eta|<2.4$ are collected in the neighbouring even and odd bins in multiplicity 
(\ref{eq:r11}) and multiplied by the powers of 0.2 for different $\eta_c$. 
The data confirm existence of the complex structure observed in MD by the 
E735 and UA5 Collaborations at energies 
$\sqrt{s}=900$-1800~GeV in the full phase-space. 
At $\sqrt{s}=7000$~GeV,
the peaky form at low $n$ and a shoulder at high multiplicities 
is seen in the all five measured $\eta_c$-windows. The peak becomes most prominent
and the shoulder most wide for $|\eta|<2.4$. 
The evolution of the structure with $\eta_c$ is demonstrated in Fig. 3(b) in more detail where the relative residues of the data with respect to the NBD parametrization are depicted.
Fore the sake of clarity the residues are mutually shifted by unity for
different pseudorapidity intervals.
The solid lines represent a description of the data by GHD.  
One can see from Fig.3(b) that at the energy $\sqrt{s}=7000$~GeV  
the residual structure survives even down to $\eta_c=0.5$.
As a consequence, NBD is not sufficient to describe the data well 
enough neither for small pseudorapidity windows at this super high energy.    
%
%
\label{Figure:3}
\begin{figure}[b] 
\vskip 3.2cm
\begin{center}
\hspace*{-1.5cm}
\parbox{6cm}{\epsfxsize=5.cm\epsfysize=4.3cm\epsfbox[95 95 400 400]
{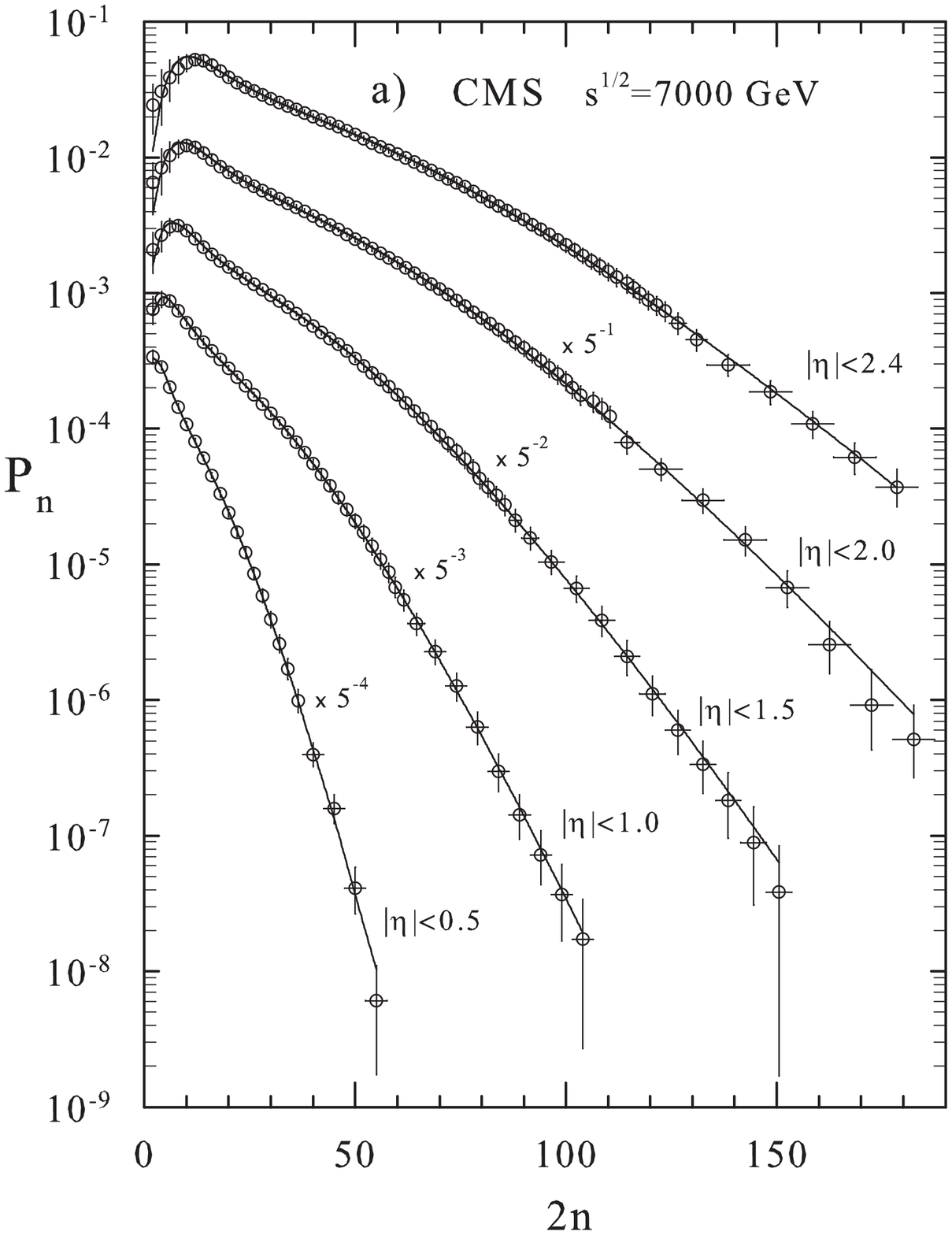}{}}
\hspace*{2.5cm}
\parbox{6cm}{\epsfxsize=5.cm\epsfysize=4.3cm\epsfbox[95 95 400 400]
{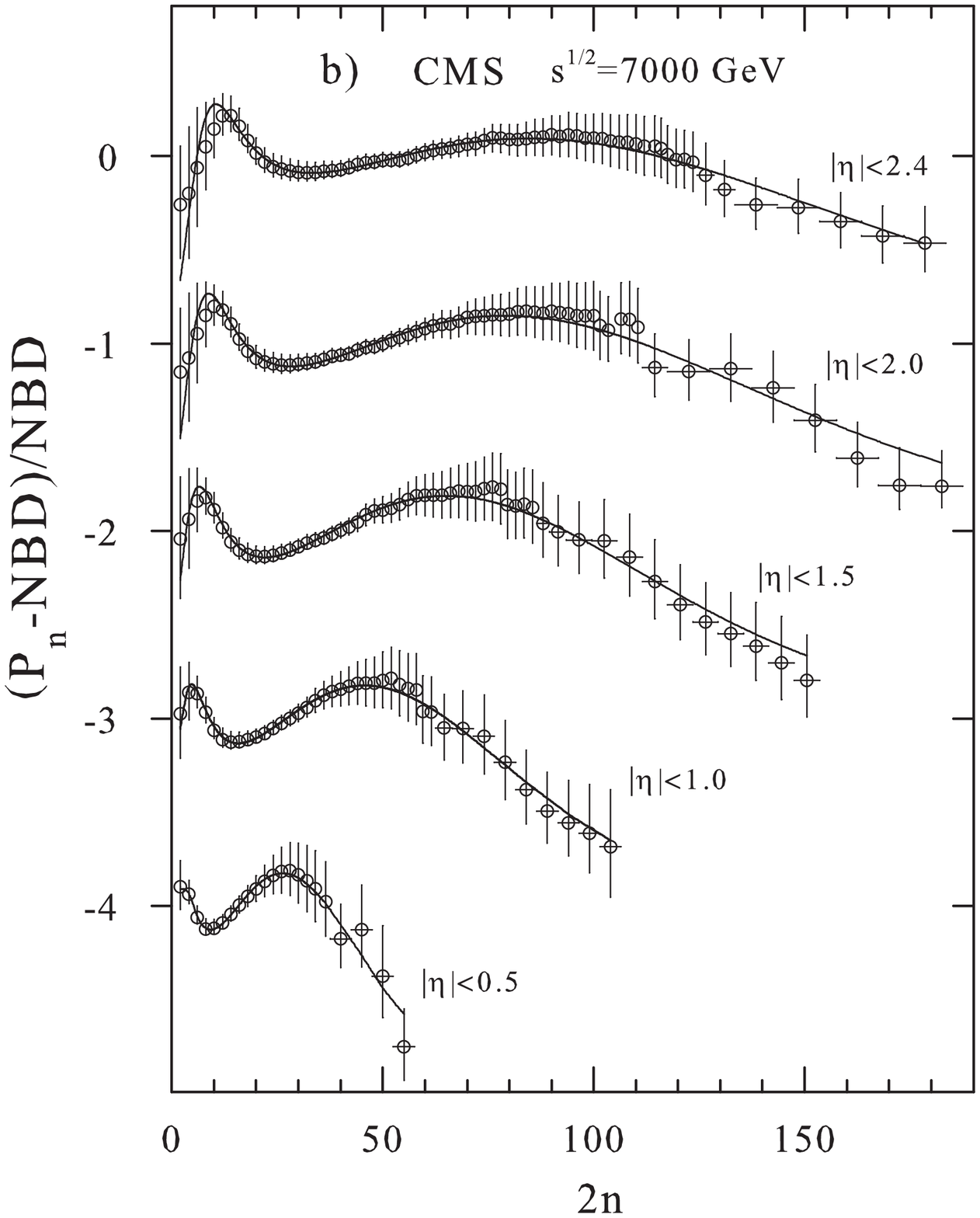}{}}
\vskip 1.3cm
\end{center}
\caption{
(a) MD of charged particles in the limited phase-space regions 
from the NSD $pp$ collisions at $\sqrt{s}=7000$~GeV. 
The data \cite{CMS} are accumulated in the neighbouring even and odd 
bins and multiplied by the indicated factors.
(b) The corresponding relative residues with respect to the NBD parametrization.
The residues are shifted mutually by unity for different $\eta_c$. 
The lines represent description of the data with GHD.}
\end{figure}
%
%
\label{Figure:4}
\begin{figure}[t] 
\vskip 3.4cm
\begin{center}
\hspace*{-1.5cm}
\parbox{6cm}{\epsfxsize=5.cm\epsfysize=4.3cm\epsfbox[95 95 400 400]
{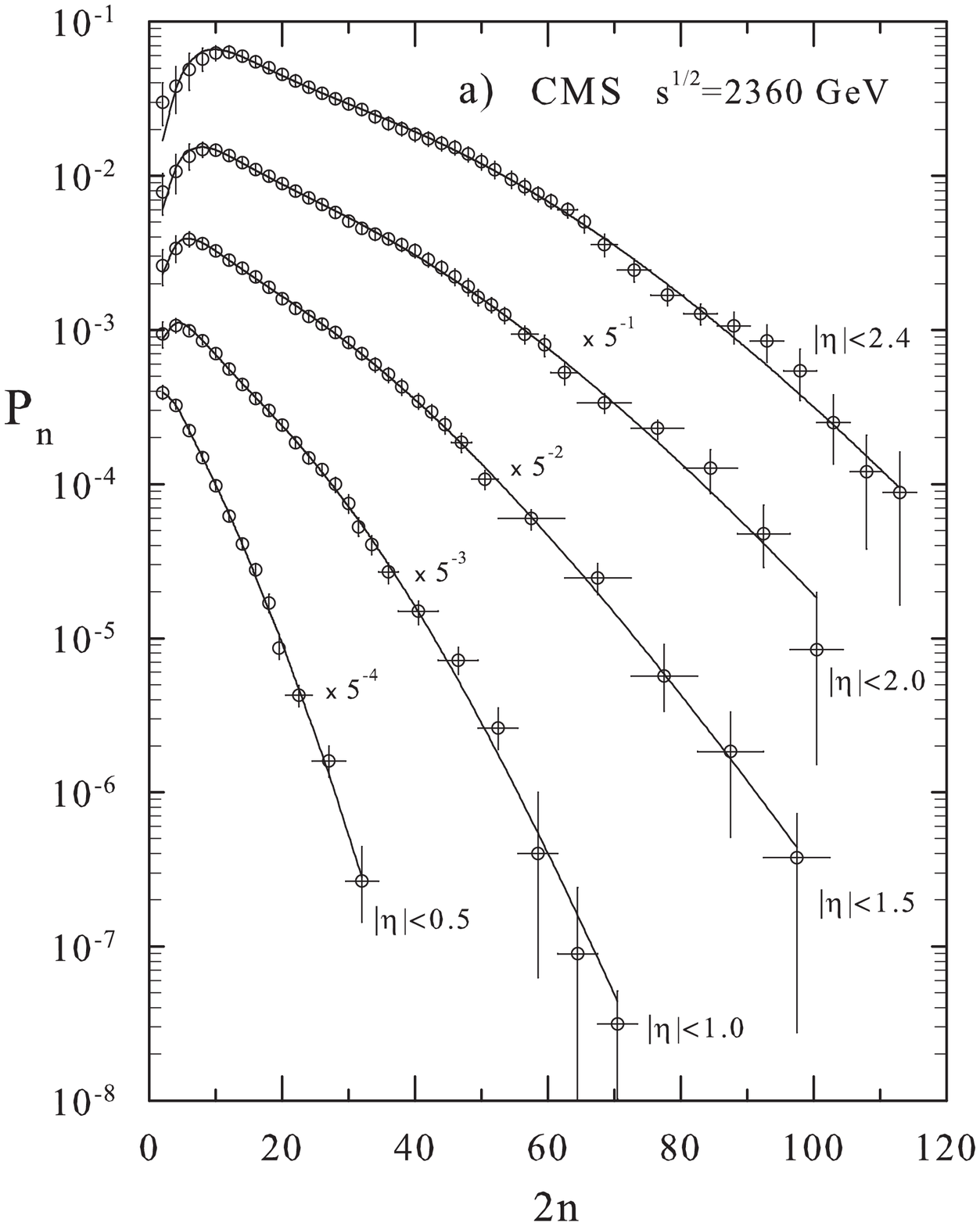}{}}
\hspace*{2.5cm}
\parbox{6cm}{\epsfxsize=5.cm\epsfysize=4.3cm\epsfbox[95 95 400 400]
{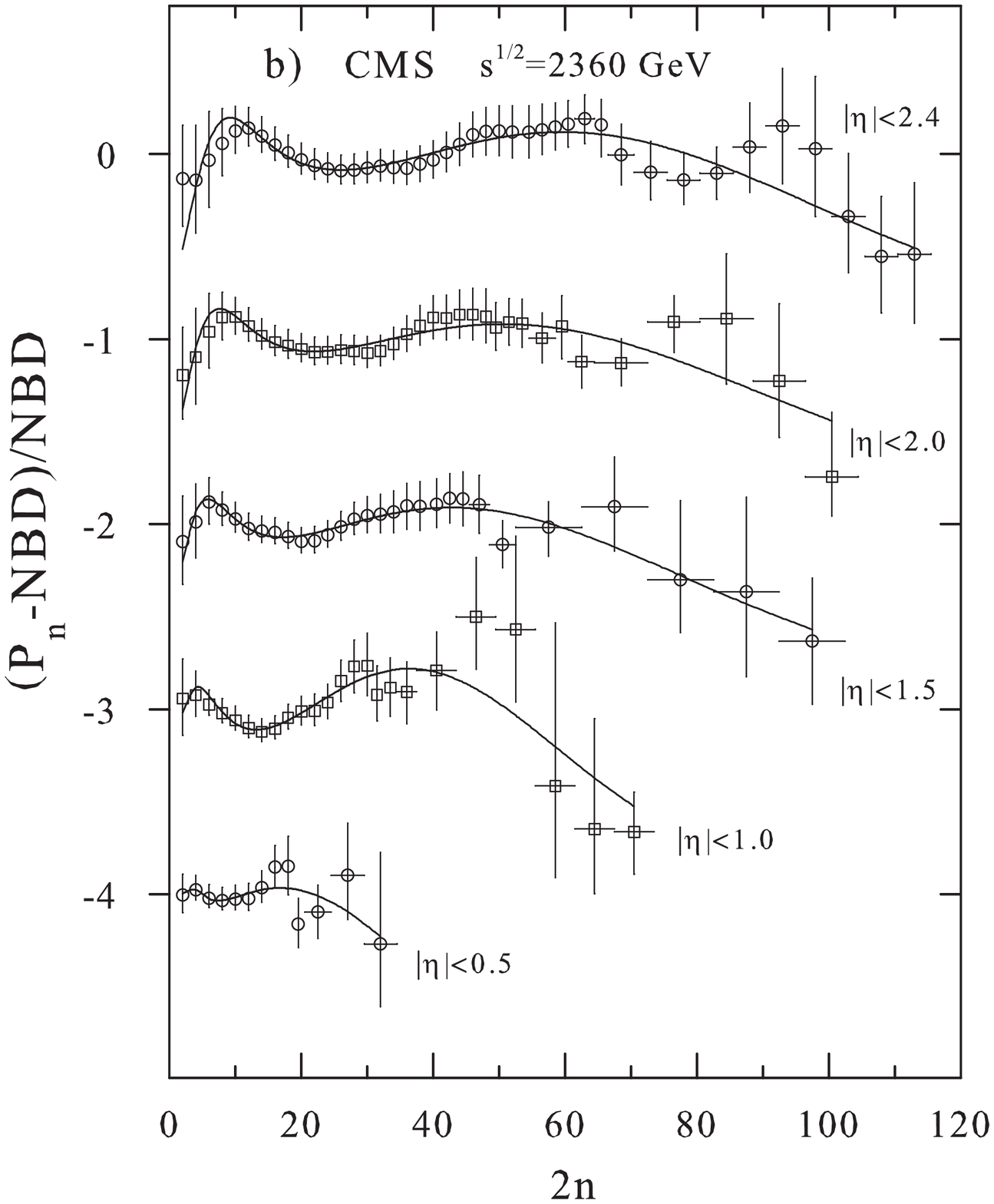}{}}
\vskip 1.3cm
\end{center}
\caption{
(a) MD of charged particles in the limited phase-space regions 
from the NSD $pp$ collisions at $\sqrt{s}=2360$~GeV. 
The data \cite{CMS} are accumulated in the neighbouring even and odd 
bins and multiplied by the indicated factors.
(b) The corresponding relative residues with respect to the NBD parametrization.
The residues are shifted by unity for different $\eta_c$. 
The lines represent description of the data with GHD. }
\end{figure}
%
%
\label{Figure:5}
\begin{figure}[h!] 
\vskip 3.8cm
\begin{center}
\hspace*{-1.5cm}
\parbox{6cm}{\epsfxsize=5.cm\epsfysize=4.3cm\epsfbox[95 95 400 400]
{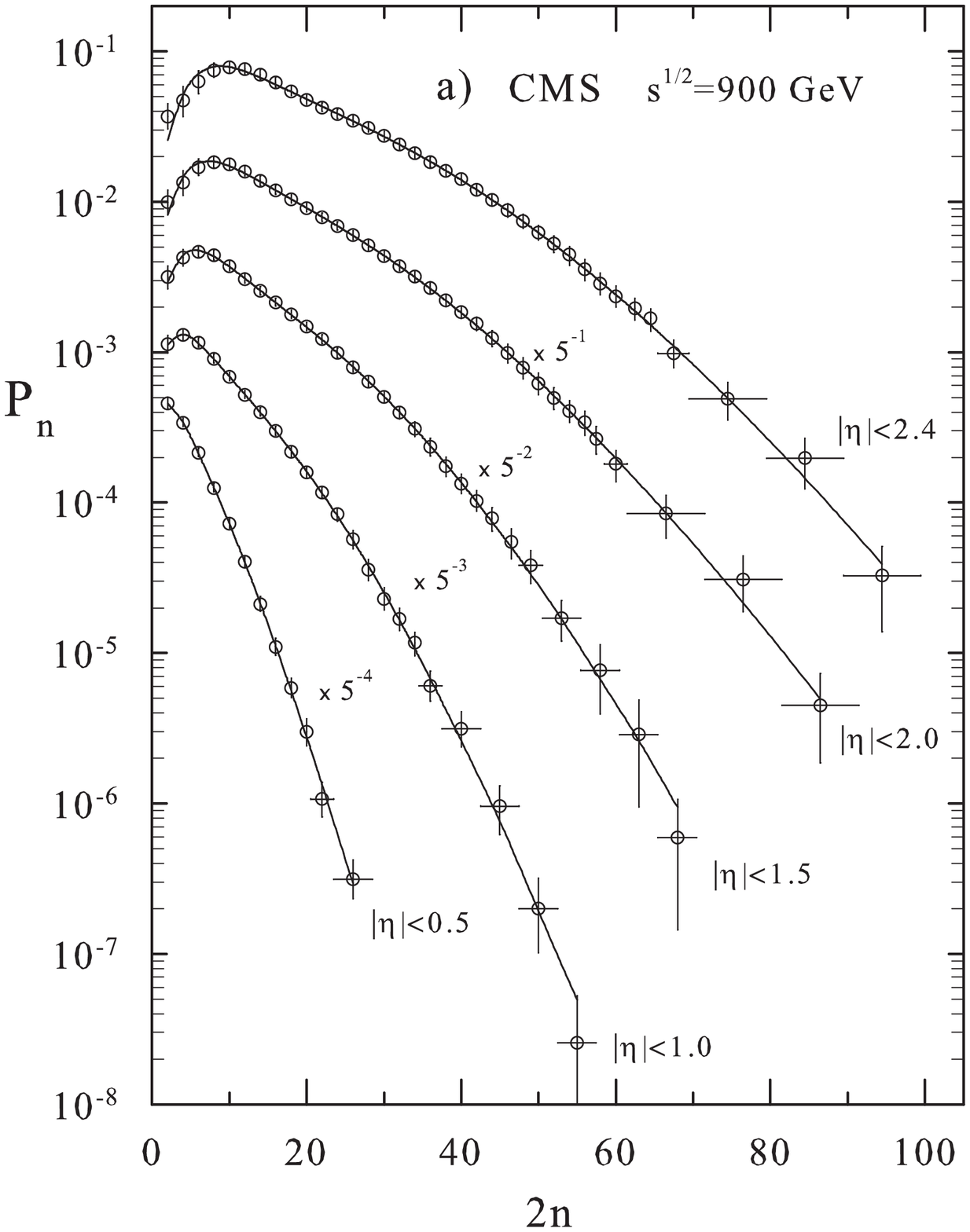}{}}
\hspace*{2.5cm}
\parbox{6cm}{\epsfxsize=5.cm\epsfysize=4.3cm\epsfbox[95 95 400 400]
{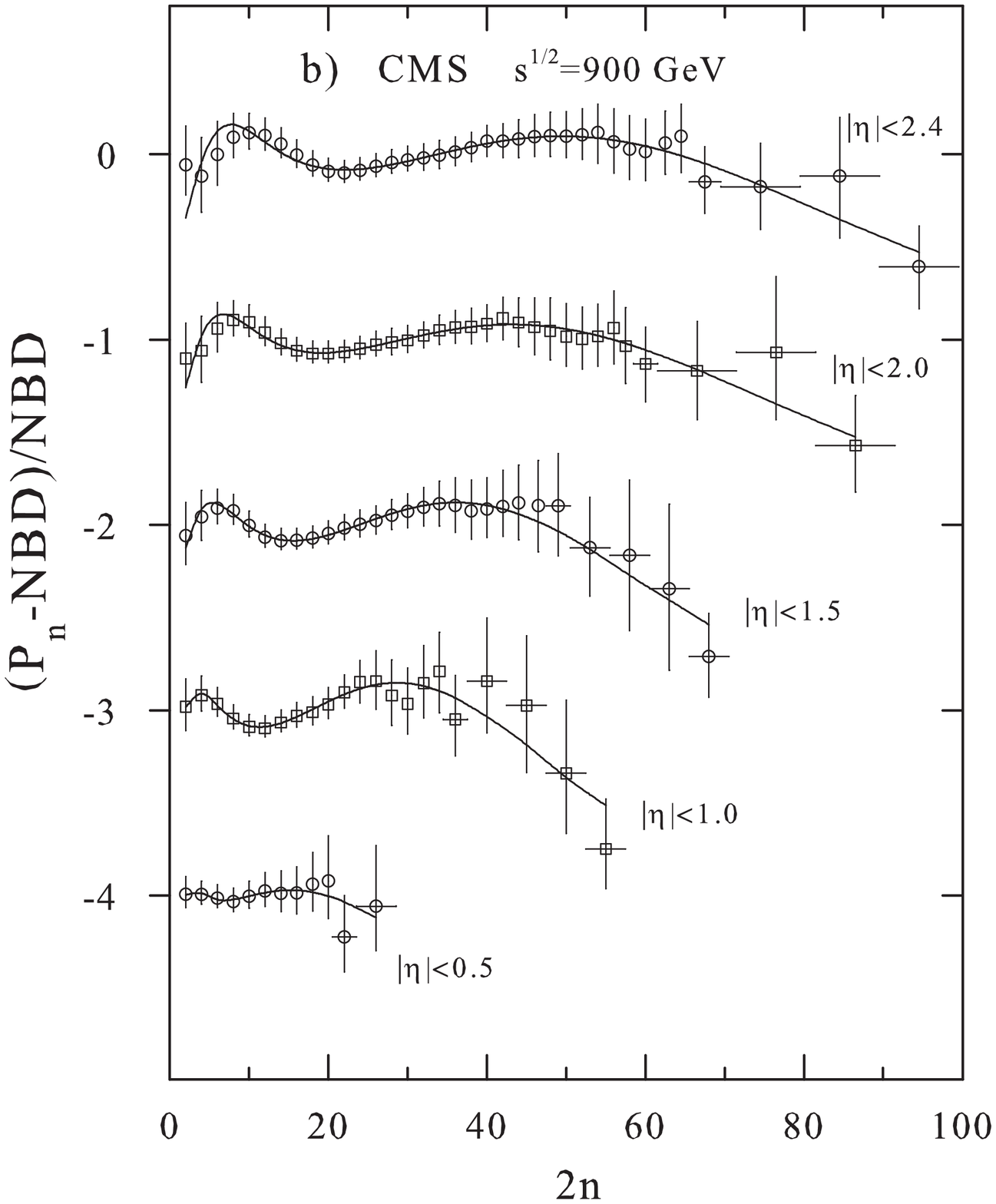}{}}
\vskip 1.5cm
\end{center}
\caption{
(a) MD of charged particles in the limited phase-space regions 
from the NSD $pp$ collisions at $\sqrt{s}=900$~GeV. 
The data \cite{CMS} are accumulated in the neighbouring even and odd 
bins and multiplied by the indicated factors.
(b) The corresponding relative residues with respect to the NBD parametrization.
The residues are shifted by unity for different $\eta_c$. 
The lines represent description of the data with GHD.}
\end{figure} 

The CMS data  measured at the energies $\sqrt{s}=2360$~GeV  and $\sqrt{s}=900$~GeV 
are presented in the same fashion in Fig. 4 and Fig. 5, respectively. 
A similar structure as in Fig. 3 is visible also at these energies.  
The systematic measurements of the CMS Collaboration allow us to study the evolution  
of the observed structure with $\sqrt{s}$.  
One can see from Figs. 3(b), 4(b), and 5(b) that the peak at low $n$ becomes 
more distinct with the increasing energy.
%
%
\label{Figure:6}
\begin{figure}[h!] 
\vskip 3.8cm
\begin{center}
\hspace*{-1.5cm}
\parbox{6cm}{\epsfxsize=5.cm\epsfysize=4.3cm\epsfbox[95 95 400 400]
{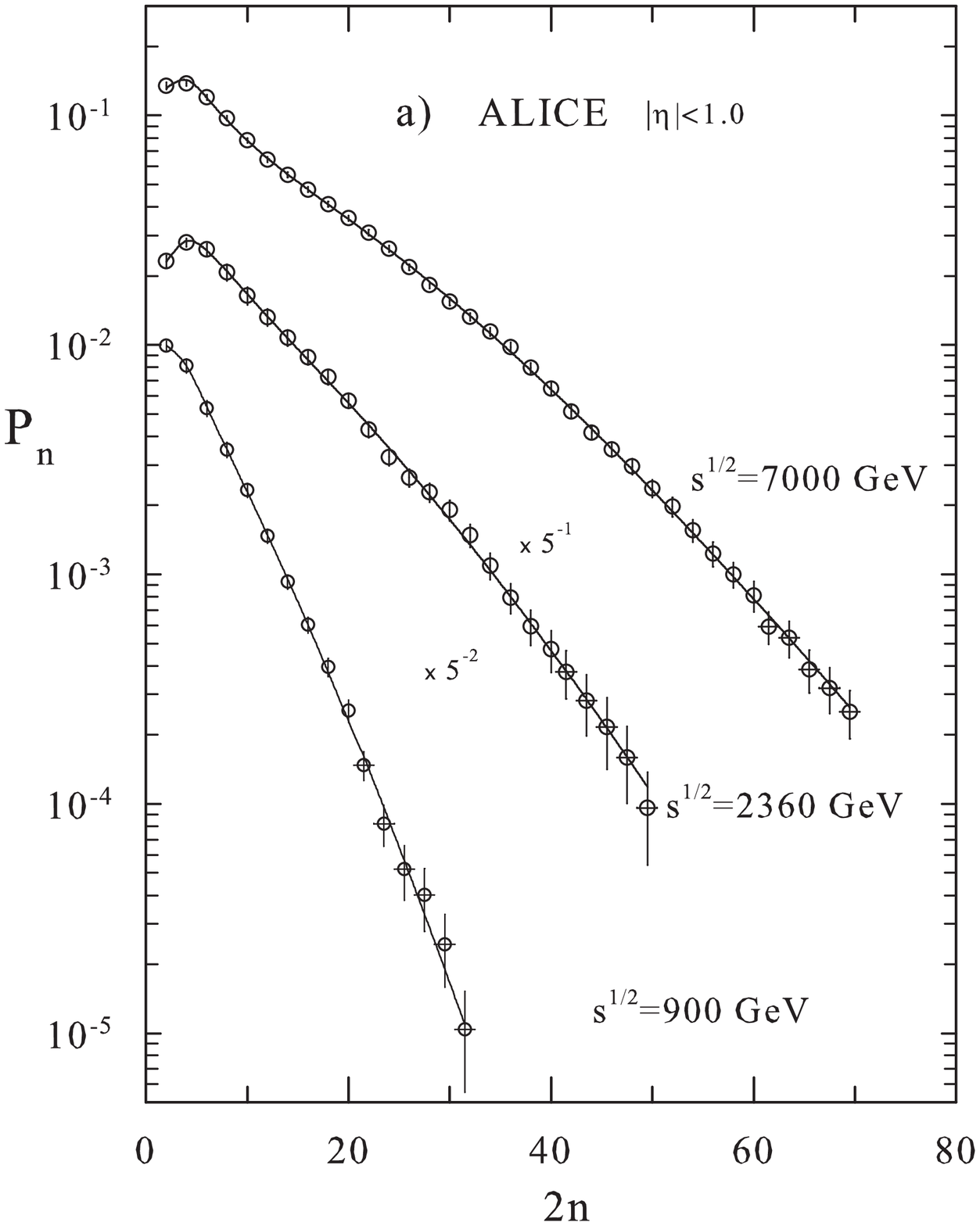}{}}
\hspace*{2.5cm}
\parbox{6cm}{\epsfxsize=5.cm\epsfysize=4.3cm\epsfbox[95 95 400 400]
{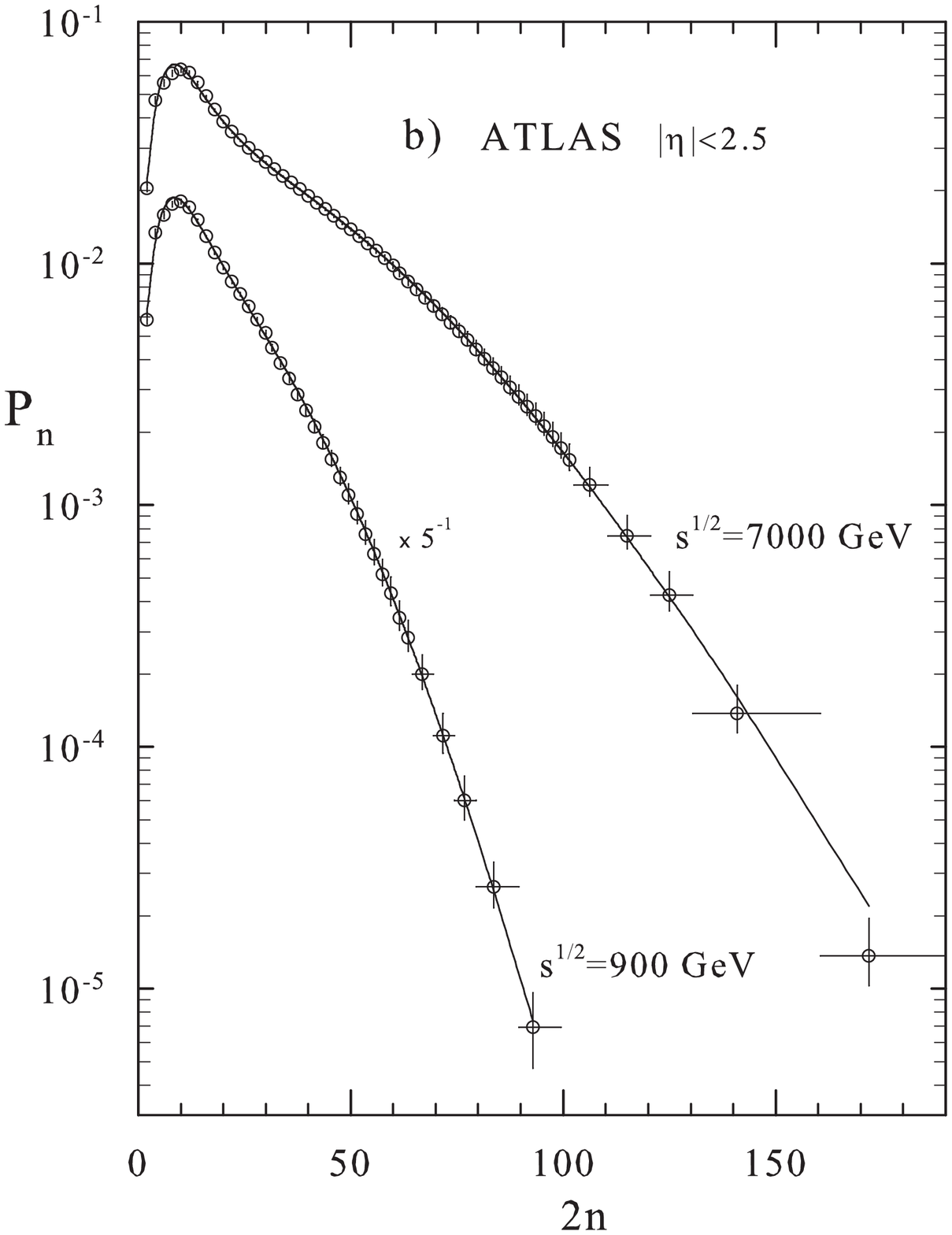}{}}
\vskip 1.5cm
\end{center}
\caption{
(a) MD of charged particles obtained by the ALICE Collaboration \cite{ALICE2} 
at the energies $\sqrt{s}=900$, 2360, and 7000~GeV 
in the central pseudorapidity region $|\eta|<1$ in the sample $n_{ch}> 0$.
(b) MD of charged particles measured by the ATLAS Collaboration \cite{ATLAS} 
at the energies $\sqrt{s}=900$ and 7000~GeV 
in the pseudorapidity region $|\eta|<2.5$ with the selection $p_T > 100$~MeV and  
$n_{ch}\ge 2$.
Data are accumulated in the neighbouring even and odd bins and 
multiplied by the indicated factors.
The lines represent description of the data with GHD. }
\end{figure}
%
%
\label{Figure:7}
\begin{figure}[h!] 
\vskip 1.8cm
\begin{center}
\hspace*{-1.5cm}
\parbox{6cm}{\epsfxsize=5.cm\epsfysize=4.3cm\epsfbox[95 95 400 400]
{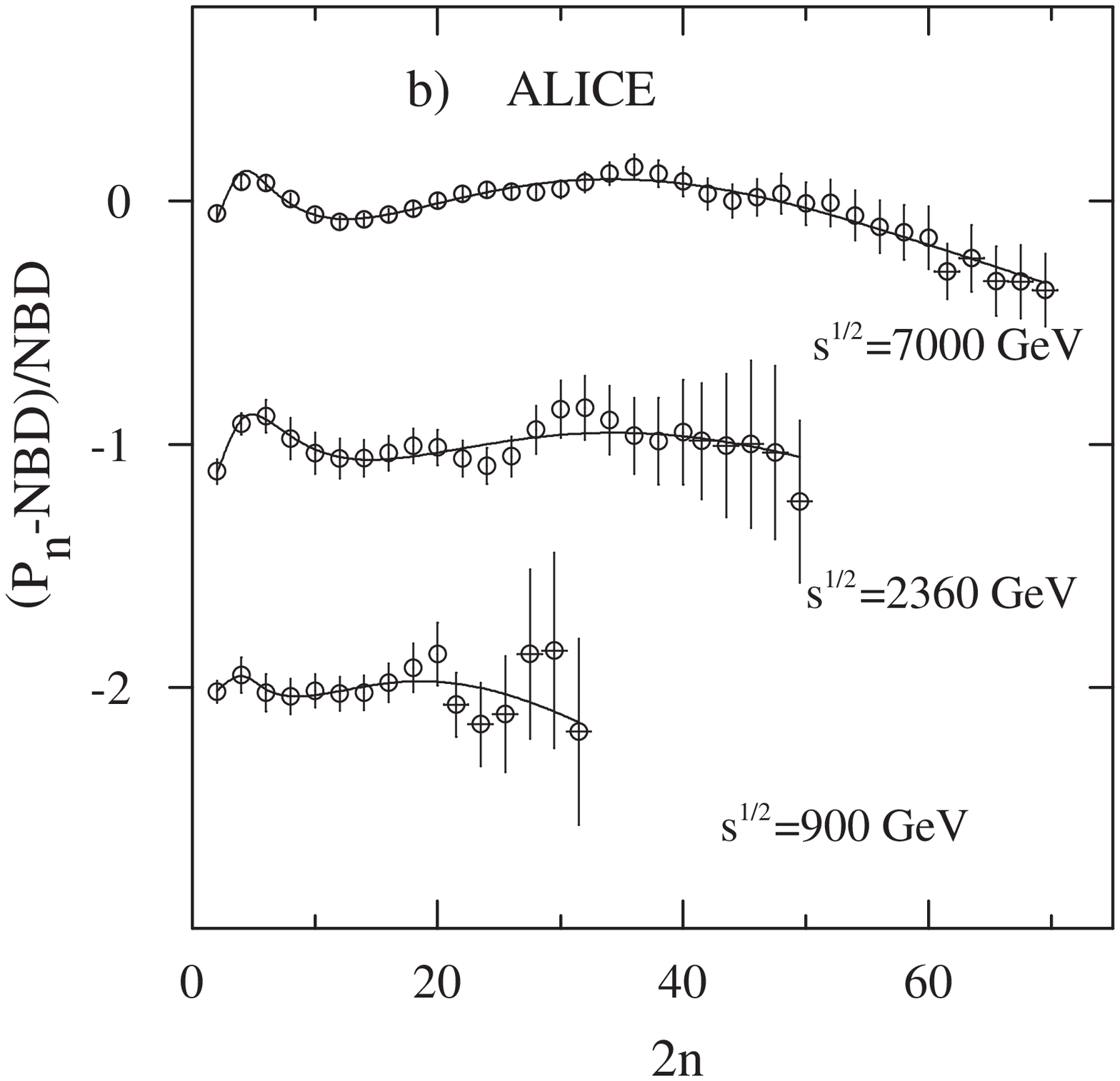}{}}
\hspace*{2.5cm}
\parbox{6cm}{\epsfxsize=5.cm\epsfysize=4.3cm\epsfbox[95 95 400 400]
{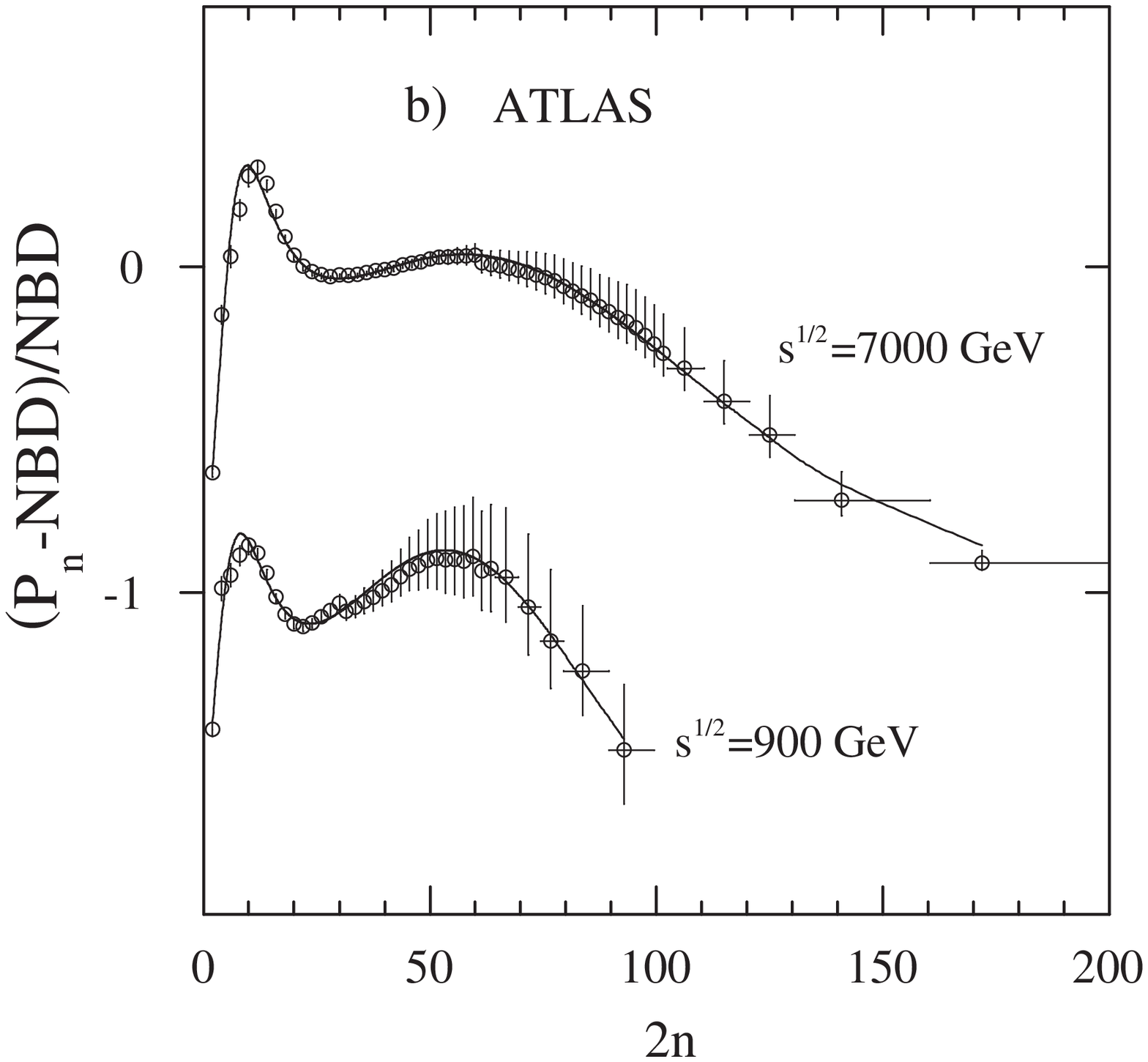}{}}
\vskip 1.2cm
\end{center}
\caption{
(a) The relative residues of the data from Fig.6(a) with respect to the NBD parametrization.
(b) The relative residues of the data from Fig.6(b) with respect to the NBD parametrization.
The residues are shifted by unity for different $\sqrt{s}$. 
The lines represent description of the data with GHD. }
\end{figure}
 
At the same time the shoulder widens 
and its maximum moves towards larger multiplicity.   
On the other hand one can see 
form Figs. 4(b) and 5(b) that, for the smallest window $|\eta|<0.5$, 
the multiplicity structure vanishes and the residues of MD with respect to the NBD 
parametrization become flat at both these energies.
This is in accord with the conclusions of Ref. \cite{Mizoguchi}, namely that 
at $\sqrt{s}=900$ and 2360~GeV the experimentally measured multiplicity distributions  
are well described by NBD for $\eta_c=0.5$.  

The ALICE Collaboration presented data \cite{ALICE2} on MD of charged particles 
in  $pp$ collisions at the energies $\sqrt{s}=900$, 2360, and 7000~GeV. 
Figure 6(a) shows the experimentally measured MD collected in the neighbouring even 
and odd bins in multiplicity (\ref{eq:r11}) in the central pseudorapidity region $|\eta|<1$.   
The data are from an event class where at least one 
charged particle in the measured pseudorapidity range is required.
The depicted distributions are multiplied by the powers of 0.2 for different 
$\sqrt{s}$. 
The corresponding relative residues with respect to the NBD parametrization 
are presented in Fig. 7(a).
The residues are mutually shifted by unity for different energies.
The full lines represent description of the ALICE data by GHD. 
One can see from Fig. 7(a) that data on MD for $|\eta|<1$ at the energy $\sqrt{s}=900$~GeV 
can be well described by NBD. 
At $\sqrt{s}=2360$~GeV, NBD is still good description of data with 
the exception at low $n$ where a peaky structure begins to emerge. 
The peak at low multiplicity is clearly visible especially at $\sqrt{s}=7000$~GeV. 
One can see from Fig. 7(a) that NBD overestimates experimental data for high 
multiplicities $(n>55)$ at this super high energy, as was already observed in Ref. \cite{ALICE2}.

The ATLAS Collaboration measured the charged-particle MD \cite{ATLAS} in different 
phase-space regions for various multiplicity cuts at three LHC energies. 
The most inclusive  phase-space region covered by the measurements corresponds to
the conditions $|\eta|<2.5$, $p_T > 100$~MeV and  $n_{ch}\ge 2$.
We analyse the ATLAS data obtained under the above $\eta_c$ and $p_T$ selections at 
$\sqrt{s}=900$ and 7000~GeV.
Figure 6(b) shows the experimentally measured MD collected in the neighbouring even 
and odd bins in multiplicity (\ref{eq:r11}).  
%
%
\label{Figure:8}
\begin{figure}[b] 
\vskip 3.8cm
\begin{center}
\hspace*{-1.5cm}
\parbox{6cm}{\epsfxsize=5.cm\epsfysize=4.3cm\epsfbox[95 95 400 400]
{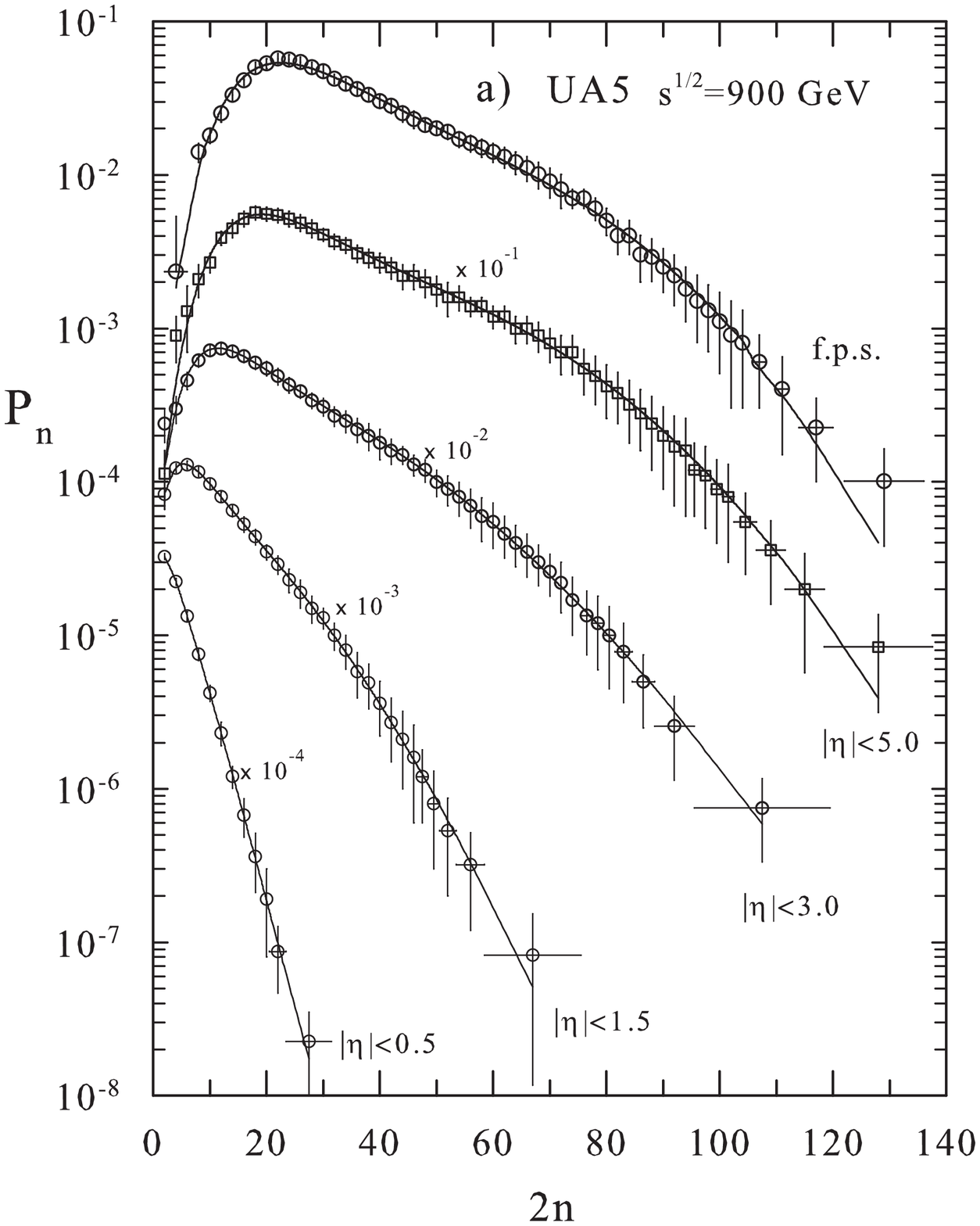}{}}
\hspace*{2.5cm}
\parbox{6cm}{\epsfxsize=5.cm\epsfysize=4.3cm\epsfbox[95 95 400 400]
{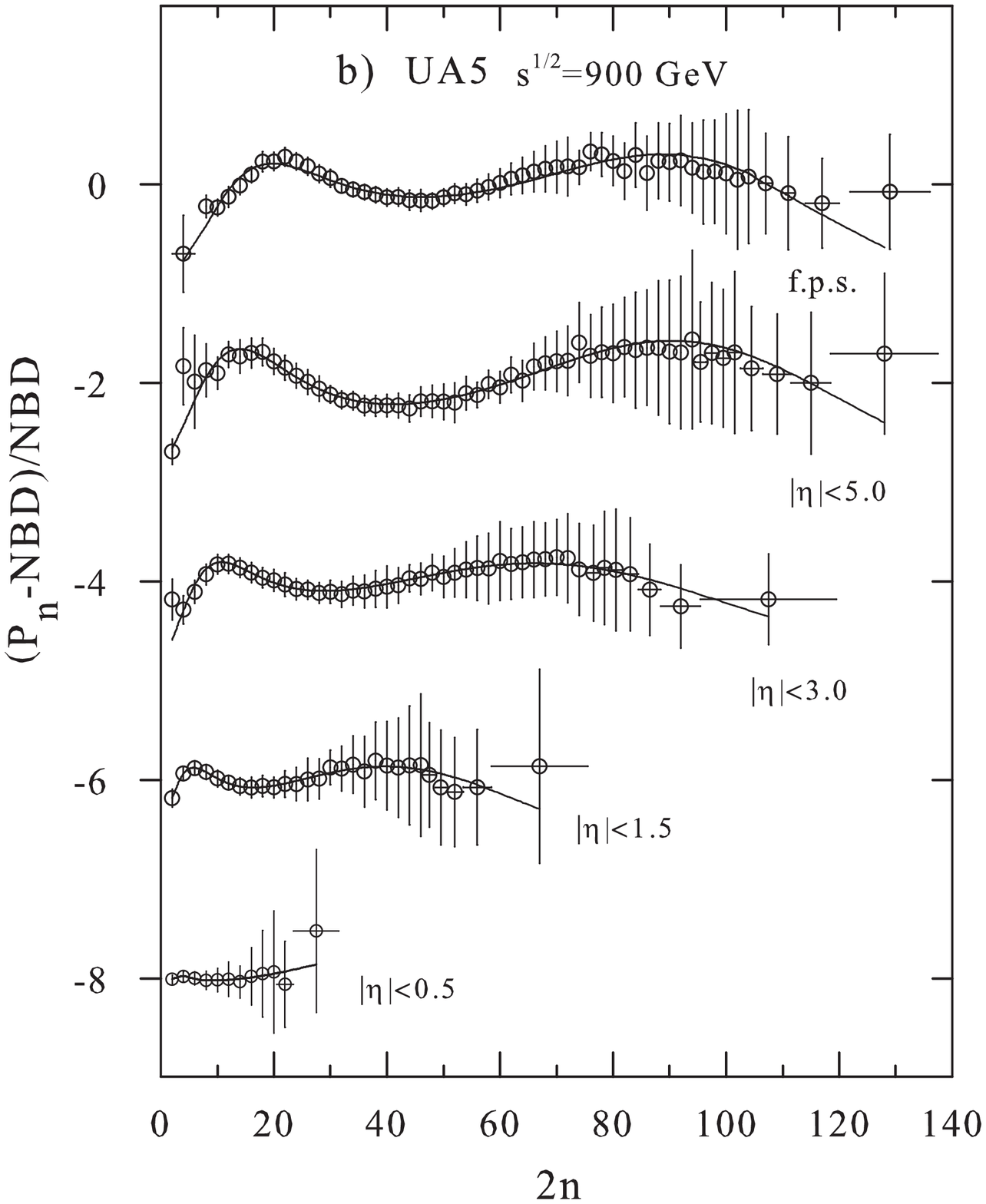}{}}
\vskip 1.5cm
\end{center}
\caption{
(a) MD of charged particles in the full phase-space and in the limited phase-space regions 
from the NSD $p\bar{p}$ collisions at $\sqrt{s}=900$~GeV. 
The data \cite{UA5} are accumulated in the neighbouring even and odd bins 
and multiplied by the indicated factors.
(b) The corresponding relative residues with respect to the NBD parametrization.
The residues are shifted mutually by the factor of two for different $\eta_c$. 
The lines represent description of the data by GHD. }
\end{figure}
The distribution at $\sqrt{s}=900$~GeV is multiplied by the factor of 0.2. 
Figure 7(b) shows the corresponding  relative residues with respect to the NBD parametrization.
The residues at $\sqrt{s}=900$~GeV are shifted down by the factor of unity. 
The full lines represent approximation of the ATLAS data by GHD. 
The distinct peak at low multiplicities in the ATLAS data represent stringent criterion 
for description of the whole shape of the distribution. 
Values of the parameters and processes included in GHD  
are strongly restricted by the peak's position, its width and hight as well as 
by the form of the shoulder at high $n$.
The four parametric GHD accounts for smooth transition from the peak region to the shoulder
at high multiplicities. Both structures in the ATLAS data are measured at such level of 
the experimental errors which exclude other processes to be present in GHD but  
those considered.
Beside this, we have checked that application of GHD to the MD of particle pairs is substantial because when applied to the distribution of all charged particles, the form of data is 
not reproduced correctly.    
The measurements performed by the CMS, ALICE, and ATLAS Collaborations show that 
the peaky structure of MD is clearly seen in $pp$ collisions            
in the limited phase-space regions at the LHC energies. 
The structure becomes more distinct with the increasing width of the pseudorapidity window
and is most pronounced at the highest energy $\sqrt{s}=7000$~GeV.

A similar trend in the pseudorapidity dependence of MD is visible 
in data \cite{UA5} measured by the UA5 Collaboration 
in the NSD $p\bar{p}$ collisions at $\sqrt{s}=900$~GeV.
Figure 8(a) shows the UA5 data collected in the neighbouring even 
and odd bins in multiplicity (\ref{eq:r11}) 
in the full phase-space and in four smaller pseudorapidity regions. 
The depicted distributions are multiplied by the powers of 0.1 for different 
$\eta_c$. 
Figure 8(b) shows the corresponding  relative residues with respect to the NBD parametrization. 
The residues in the smaller pseudorapidity windows are mutually shifted down 
by the factor of two. 
The full lines represent description of the UA5 data by GHD. 
Starting from the window $|\eta|<1.5$ a small  
peak at low multiplicity emerges in the shape of MD. It evolves with the window size and becomes 
best visible in the full phase-space. As can be seen from Fig. 8(b), the residues 
with respect to NBD  are perfectly flat in the smallest window $|\eta|<0.5$
meaning that NBD describes MD in this region well.

\subsection{Multiplicity distribution in $e^{+}e^{-}$ annihilations}

The study of multihadron production in $e^{+}e^{-}$ annihilations 
can provide additional information on correlation structures which are  
encoded in MD in an integrated form.  
The qualitative picture of MD of charged hadrons in leptonic processes  
differs in many properties form $pp/p\bar{p}$ interactions. 
The average multiplicities are higher and the distributions are narrower in $e^{+}e^{-}$ annihilation in comparison with the hadron collisions at the same energies. 
At phenomenological level, a comparison of both types of the interactions 
may reveal which features reflect difference in the initial state and which are 
intrinsic to the parton-hadron transitions.

Using formula (\ref{eq:r10}) for GHD, we studied the inclusive samples of MD of 
charged hadrons produced 
in $e^{+}e^{-}$ annihilations in the full phase-space at various collision energies 
$\sqrt{s}$. Similarly as in the previous section, 
GHD is applied to the distribution of 
particle pairs for all multiplicities $n\geq n_0$. 
The value of $n_0$ is the minimal number of the charged particle pairs 
measured in experiment.  
The results of the analysis are presented in Table II and Figs. 9-10. 
We have analysed data on MD of charged particles  
obtained by the OPAL Collaboration at the centre-of-mass energies 
of $\sqrt{s}=172$, 183, and 189~GeV \cite{OPAL1}. 
The data with high statistical precision correspond to the energy region 
sufficiently far beyond $Z^{0}$ peak. 
The three data samples are measured in the multiplicity range which begins with 
$n_{ch}=8$ i.e. with $n_0=4$ particle pairs.
The analysis includes also data on MD measured by the OPAL Collaboration 
at lower energies, $\sqrt{s}=161$~GeV \cite{OPAL2}, 
$\sqrt{s}=133$~GeV \cite{OPAL3}, and at the energy of $Z^{0}$ peak, 
$\sqrt{s}=91$~GeV \cite{OPAL4}. 
The application of GHD to the OPAL data shows that the parameter $\beta_0/\beta_2$ 
is compatible with zero and, therefore, it was set to null in this high energy region.  
In such a case, GHD becomes three parameter distribution for which the recurrence 
relation  
(\ref{eq:r1}) takes the form   
\begin{equation}
\frac{(n+1)P_{n+1}}{P_n} =
\frac{\bar{\alpha}_0}{n(n-1)}+\bar{\alpha}_2 +\bar{\alpha}_3(n-2), \ \ \ \ 
\bar{\alpha}_i\equiv\alpha_i/\beta_2.
\label{eq:r12}
\end{equation}
In this limit the distribution reflects relatively sharp increase 
of $P_n$ at low $n\geq n_0$
typical for $e^{+}e^{-}$ annihilations at high energies 
and accounts for a "negative binomial tail" at large multiplicities.

%
%
\label{Figure:9}
\begin{figure}[h!] 
\vskip 3.2cm
\begin{center}
\hspace*{-1.5cm}
\parbox{6cm}{\epsfxsize=5.cm\epsfysize=4.3cm\epsfbox[95 95 400 400]
{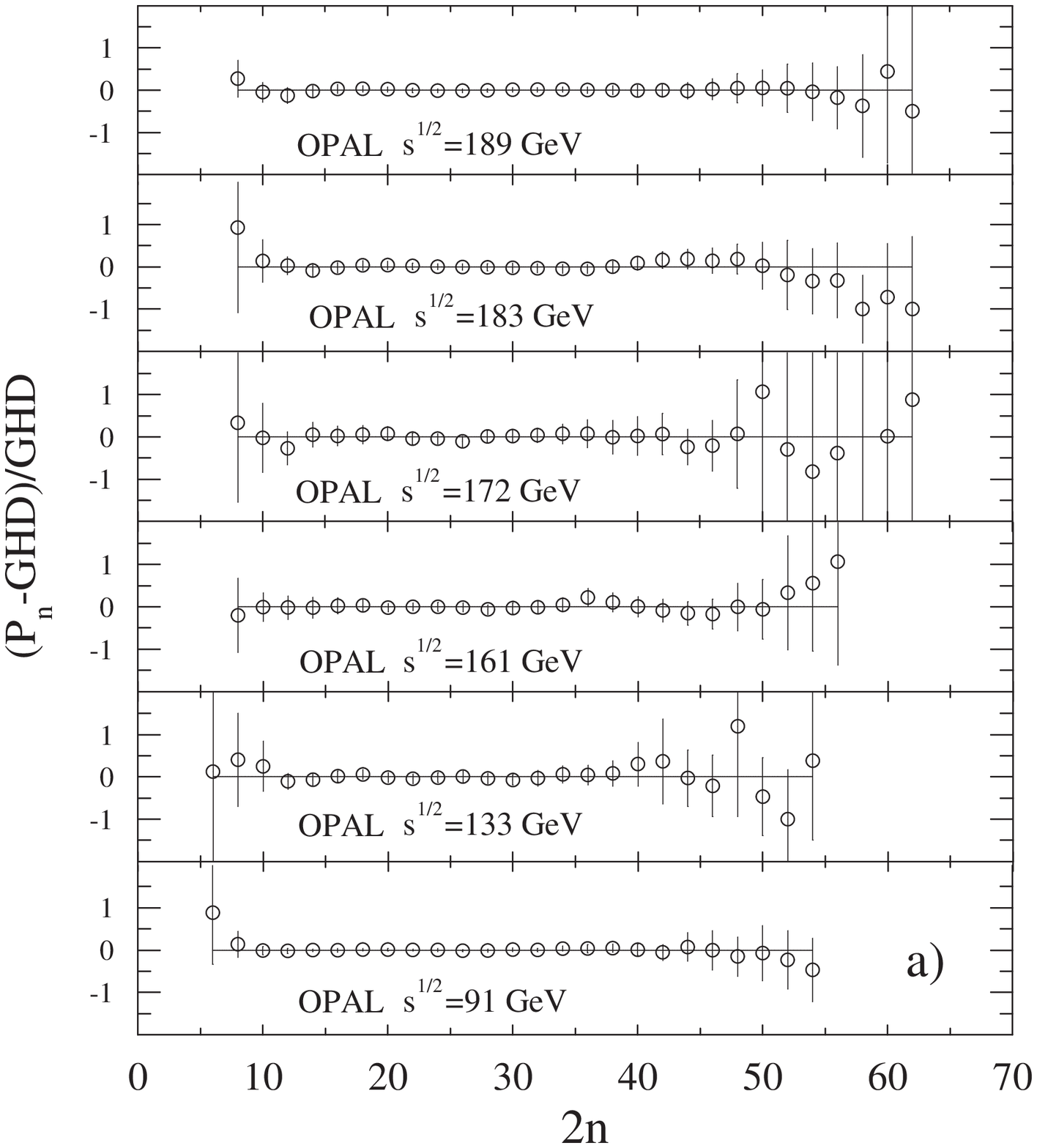}{}}
\hspace*{2.5cm}
\parbox{6cm}{\epsfxsize=5.cm\epsfysize=4.3cm\epsfbox[95 95 400 400]
{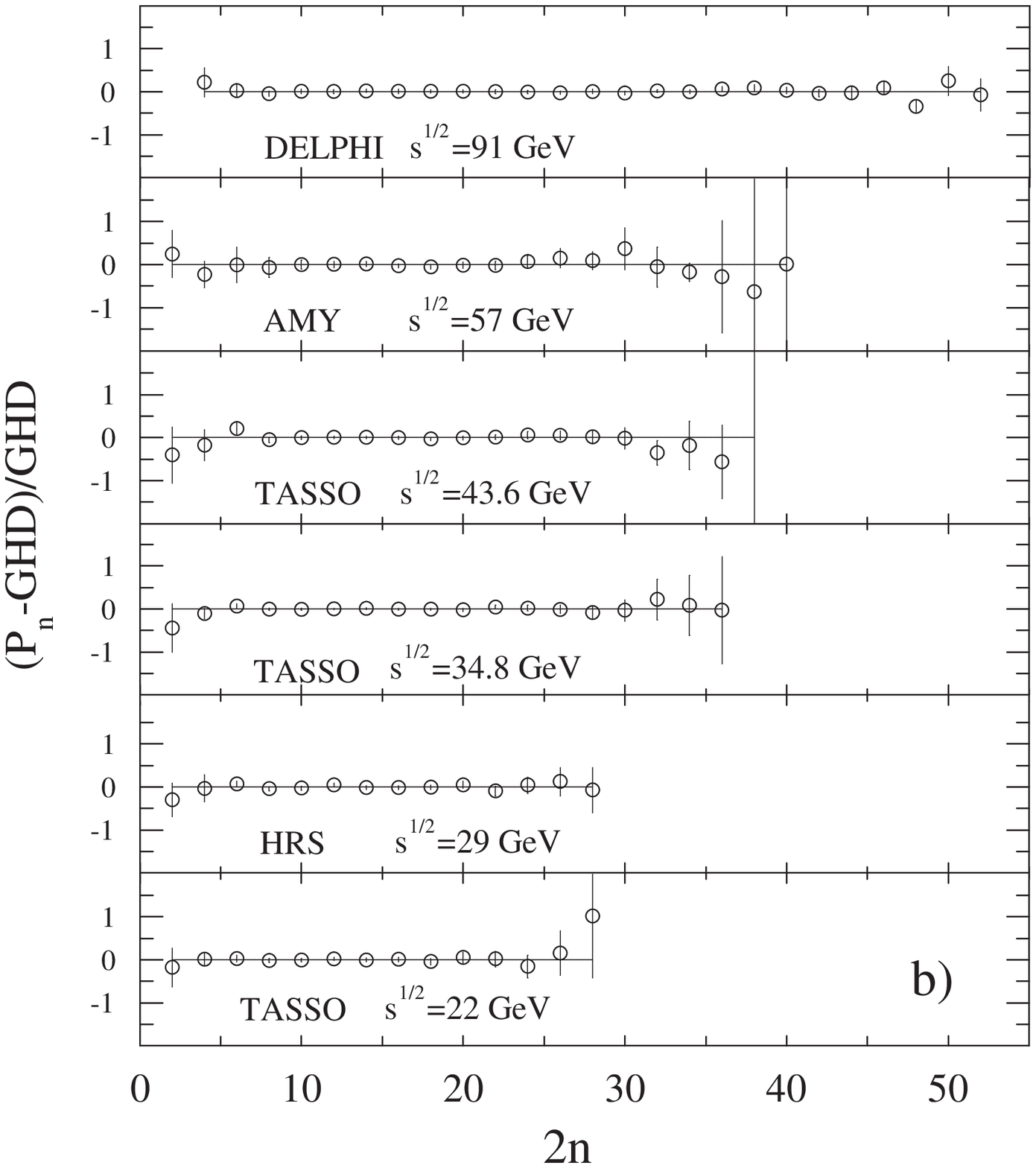}{}}
\vskip 0.2cm
\end{center}
\caption{
 The relative residues of MD of charged particles produced in 
$e^{+}e^{-}$ annihilation in the full phase-space with respect to description by GHD.
(a) The data  measured by the OPAL Collaboration 
at $\sqrt{s}=189$, 183, 172~GeV are from Ref. \cite{OPAL1}, 
at $\sqrt{s}=161$~GeV from Ref. \cite{OPAL2}, 
at $\sqrt{s}=133$~GeV from Ref. \cite{OPAL3}, and 
at $\sqrt{s}=91$~GeV from Ref. \cite{OPAL4}. 
(b) The data measured by the DELPHI Collaboration 
at $\sqrt{s}=91$~GeV are from Ref. \cite{DELPHI}. 
The combined data measured by the AMY Collaboration 
at $\sqrt{s}=57$~GeV are from Ref. \cite{AMY}. 
The data measured by the TASSO Collaboration at $\sqrt{s}=43.6$, 34.8, and 22~GeV
are from Ref. \cite{TASSO}.    
The data measured by the HRS Collaboration at $\sqrt{s}=29$~GeV
are from Ref. \cite{HRS}.} 
\end{figure}
%
%
\label{Figure:10} 
\begin{figure}[h!] 
\vskip 3.4cm
\begin{center}
\hspace*{-1.5cm}
\parbox{6cm}{\epsfxsize=5.cm\epsfysize=4.3cm\epsfbox[95 95 400 400]
{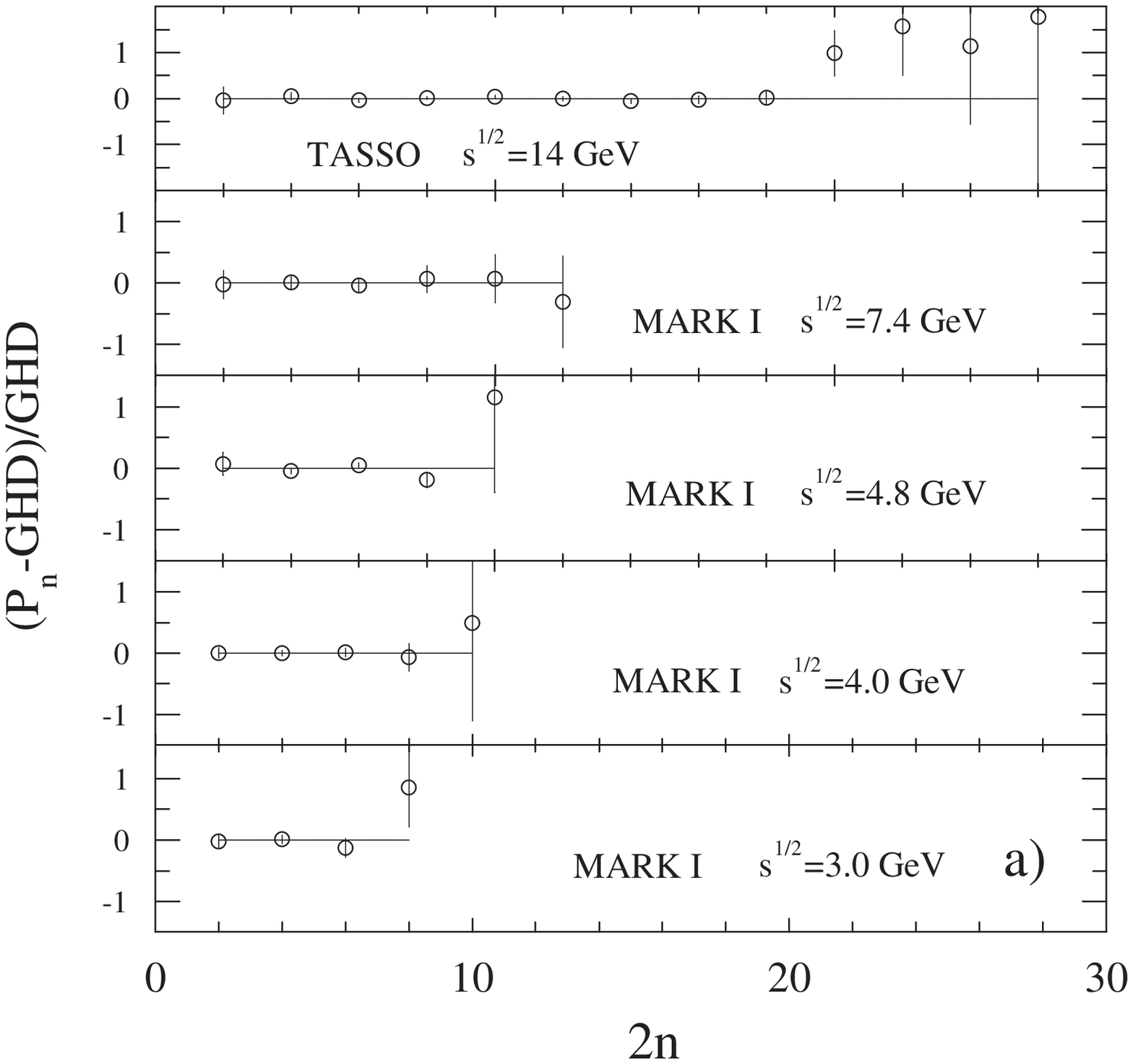}{}}
\hspace*{2.5cm}
\parbox{6cm}{\epsfxsize=5.cm\epsfysize=4.3cm\epsfbox[95 95 400 400]
{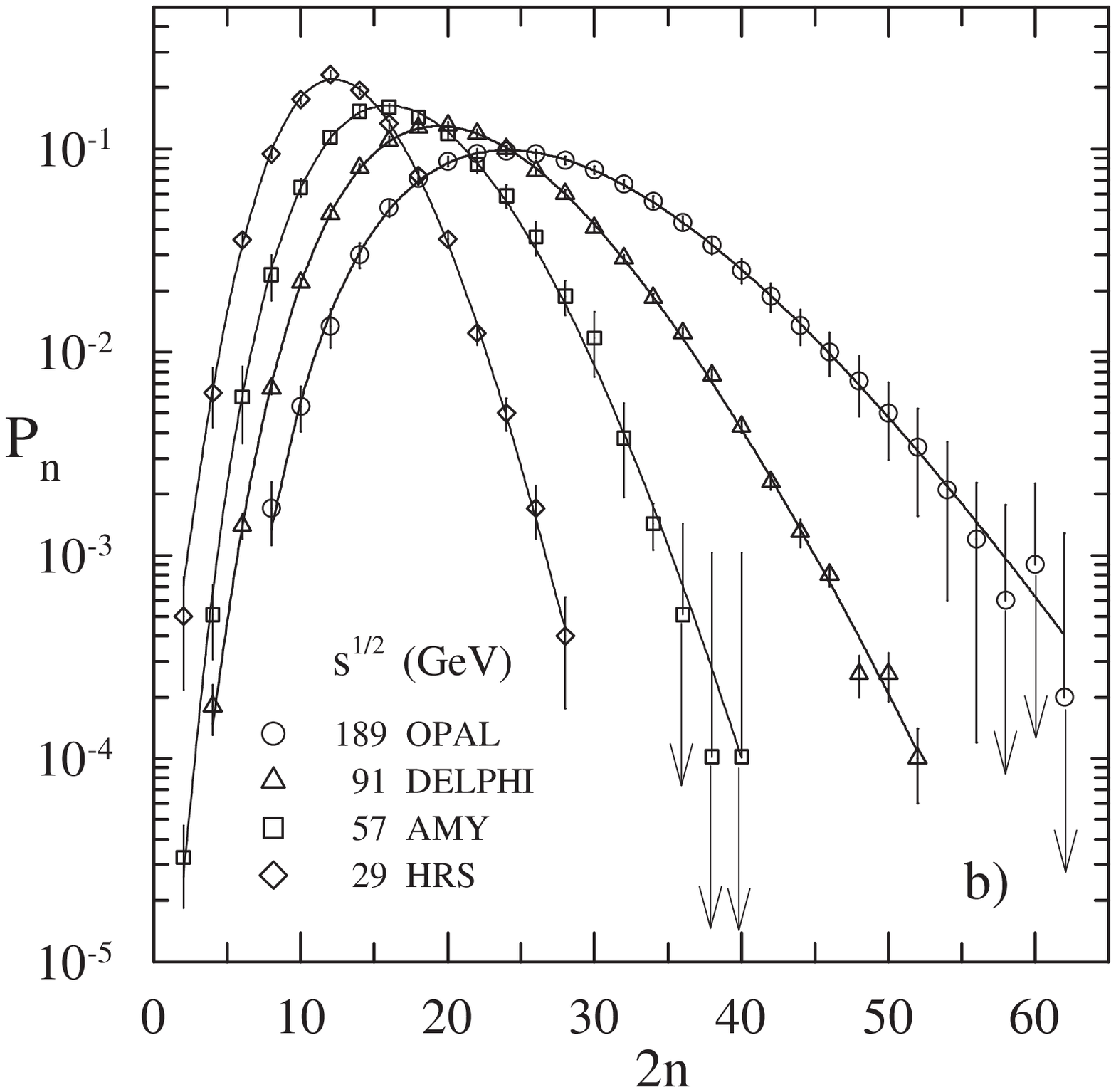}{}}
\end{center}
\vskip -1.0cm
\caption{ 
(a) The relative residues of MD of charged particles produced in 
$e^{+}e^{-}$ annihilation in the full phase-space with respect to description by GHD.
The data measured by the TASSO Collaboration 
at $\sqrt{s}=14$~GeV are from Ref. \cite{TASSO}. 
The data measured by the MARK I Collaboration 
at $\sqrt{s}=7.4$, 4.8, 4.0, and 3.0~GeV are from Ref. \cite{MARK}. 
(b) The energy dependence of MD of charged particles produced in 
$e^{+}e^{-}$ annihilation in the full phase-space. 
The depicted data from the OPAL \cite{OPAL1}, DELPHI \cite{DELPHI}, 
AMY \cite{AMY}, and HRS \cite{HRS} Collaborations were measured at the energies 
$\sqrt{s}=189$, 91, 57, and 22~GeV, respectively.   
The lines represent description of the data by GHD. }
\end{figure}

%
%
\begin{table}[htb!] 
\caption{The results of analysis of MD of the charged particle pairs in 
$e^{+}e^{-}$ annihilations in the full phase-space by GHD. 
The minimal number of the pairs is $n_0$. If not quoted, the corresponding 
parameter of GHD was set to null.
The errors correspond to the quadratic sum of the statistical and systematic uncertainties 
of data when both are published.} 
\label{tab:ee fps} 
\begin{center} 
\begin{ruledtabular} 
\begin{tabular}{ccccccccc} 
$\sqrt{s}$ & {} & {} &  {} & GHD
\\ 
\cline{4-9}
{(GeV)} & {} & {} & $n_0$ &
$\alpha_0/\beta_2$ & $\alpha_2/\beta_2$ & $\alpha_3/\beta_2$ & $\beta_0/\beta_2$ &
$\chi^2/\mbox{NDF}$ 
\\ 
\hline 
189  & OPAL  &  {} &  4  &
 $163 \pm 34$  & $6.8\pm 0.8$ & $0.47\pm 0.05$ & {-}       & 1.9/25
 \\
183  & OPAL  &  {} &  4  &
 $344 \pm 61$  & $3.0\pm 1.3$ & $0.67\pm 0.07$ & {-}       & 6.5/25 
 \\
172  & OPAL  &  {} &  4  &
 $123 \pm 108$  & $8.5\pm 2.7$ & $0.29\pm 0.17$ & {-}       & 3.0/25  
 \\
161  & OPAL  &  {} &  4  &
 $130 \pm 55$  & $6.8\pm 1.5$ & $0.40\pm 0.09$ & {-}       & 3.0/22   
 \\
133  & OPAL  &  {} &  3  &
 $198 \pm 69$  & $4.3\pm 1.8$ & $0.54\pm 0.12$ & {-}       & 3.9/22  
 \\
91  & OPAL  &  {} &  3  &
 $131 \pm 17$  & $6.1\pm 0.5$ & $0.33\pm 0.04$ & {-}       & 3.1/22   
 \\
91  & DELPHI  &  {} &  2  &
 $192 \pm 41$  & $6.2\pm 0.4$ & $0.31\pm 0.03$ & $5.4\pm 2.6$      & 11.9/21  
 \\
91  & ALEPH  &  {} &  2  &
 $127 \pm 39$  & $5.9\pm 1.4$ & $0.33\pm 0.10$ & {-}       & 3.2/23 
 \\
57  & AMY  &  {} &  1  &
 $92 \pm 33$  & $6.8\pm 1.0$ & $0.02\pm 0.09$ & $1.8\pm 1.3$      & 4.5/16     
 \\
43.6  & TASSO  &  {} &  1  &
 $141 \pm 42$  & $5.7\pm 0.2$ & {-} & $9.2\pm 4.3$      & 7.3/16      
 \\
34.8  & TASSO  &  {} &  1  &
 $191 \pm 49$  & $1.9\pm 1.1$ & $0.28\pm 0.10$ & $12.0\pm 4.1$      & 6.6/14       
 \\
29  & HRS  &  {} &  1  &
 $153 \pm 76$  & $2.7\pm 2.3$ & $0.05\pm 0.23$ & $8.4\pm 5.6$      & 5.0/10 
 \\
22  & TASSO  &  {} &  1  &
 $77 \pm 36$  & $2.6\pm 1.5$ & $0.11\pm 0.18$ & $4.3\pm 3.0$      & 2.2/10     
 \\
14  & TASSO  &  {} &  1  &
 $37 \pm 6$  & $2.3\pm 0.2$ & {-} & $2.1\pm 0.7$      & 10.3/10    
 \\
7.4  & MARK I  &  {} &  1  &
 $40 \pm 16$  & {-} & {-} & $11.9\pm 6.2$      & 0.4/4     
 \\
4.8  & MARK I  &  {} &  1  &
 $7.3\pm 0.7$  & {-} & {-} & $1.1\pm 0.3$      & 5.0/3      
 \\
4.0  & MARK I  &  {} &  1  &
 $8.1\pm 1.2$  & {-} & {-} & $2.2\pm 0.5$      & 0.2/3    
 \\
3.0  & MARK I  &  {} &  1  &
 $5.3\pm 1.8$  & {-} & {-} & $2.2\pm 0.9$      & 2.6/2      
\end{tabular} 
\end{ruledtabular} 
\end{center} 
\end{table} 
 
The relative residues of MD of charged particles measured by the OPAL Collaboration 
with respect to the description of the data by GHD with three parameters (\ref{eq:r12}) 
are depicted in Fig. 9(a). 
The residues are nearly zero reflecting acceptable parametrization of the experimental data.
Similar holds for the data \cite{ALEPH} obtained by the ALEPH Collaboration at the 
energy $\sqrt{s}=91$~GeV (see Table II).  
As shown in Ref. \cite{Dewanto}, a good description of the OPAL data can be obtained 
also by GMD characterized by the recurrence factor (\ref{eq:r5}). 
An exception represent data on MD measured by the DELPHI Collaboration 
at the energy of $Z^0$ peak in the sense that their description requires 
a non-zero value of the parameter $\beta_0/\beta_2$. 
As seen from Table II, this parameter is required also by data at lower energies. 
The relative residues of MD of charged particles measured 
by the DELPHI \cite{DELPHI}, 
AMY \cite{AMY}, TASSO \cite{TASSO}, and HRS \cite{HRS} Collaborations 
with respect to the four parametric GHD are depicted in Fig. 9(b).  
In all cases a good description is obtained.

We analysed also data \cite{MARK} on MD in $e^{+}e^{-}$ annihilations measured 
by the MARK I Collaboration in the low energy region. 
The application of GHD to the multiplicity data for $\sqrt{s}<10$~GeV gives 
negative values of the parameters $\alpha_3/\beta_2$ and $\alpha_2/\beta_2$ with 
an over-parametrized description. Therefore we set $\alpha_3=\alpha_2=0$ at these 
energies.   
In that case, similarly as for $pp$ collisions at very low energies, 
MD in $e^{+}e^{-}$ annihilation 
can be characterized by the recurrence (\ref{eq:r1}) with   
\begin{equation}
g^{-1}(n) =
\bar{\beta}_0 +\bar{\beta}_2n(n-1)  , \ \ \ \ 
\bar{\beta}_i\equiv\beta_i/\alpha_0.
\label{eq:r13}
\end{equation}
Because of the non-zero values of $\bar{\beta}_2$ in (\ref{eq:r13}), 
GHD is narrower than Poisson distribution in this region.
Figure 10(a) shows the relative residues of MD with respect 
to GHD at low energies.
A summarizing illustration of the description of MD of charged particles 
produced in $e^{+}e^{-}$ annihilations at some energies is presented in Fig. 10(b). The full lines represent GHD with parameters quoted in Table II.

\section{Discussion}

The motivation of the present study of multiple production originates from 
the experimental observation that in $pp/p\bar{p}$ interactions at high energies a structure in 
the charged particle MD emerges both in the full phase-space and in the limited phase-space regions.
We concentrate in obtaining a plausible description of the observed structure 
which is distinctly visible in new data from the LHC in the super high energy domain.  
The phenomenological analysis is based on a scenario of multiparticle production 
in terms of parton cascade processes.
The proposed approach aims to grasp some qualitative features of parton to hadron 
transitions  which may be important at the end of the parton cascading.  
The observable shape of MD is assumed to be influenced mostly by the soft particles 
produced in the final stages of the cascade development.
Besides the ordinary birth $(0\rightarrow 1)$ and death $(1\rightarrow 0)$ process, 
we consider the multiparton incremental $(2\rightarrow 3)$, $(3\rightarrow 4)$  
and decremental 
$(3\rightarrow 2)$ recombination processes which are supposed to contribute significantly 
to the multiplicity build up. 
Such kind of two and three parton recombination interactions in the final stage of 
the cascade evolution can be justified by the physical 
requirements of color neutralization and reaching of an approximate 
"momenta uniformity" at hadronization.

%
%
\label{Figure:11}
\begin{figure}[] 
\vskip 1.5cm
\begin{center}
\hspace*{-1.5cm}
\parbox{6cm}{\epsfxsize=5.cm\epsfysize=4.3cm\epsfbox[95 95 400 400]
{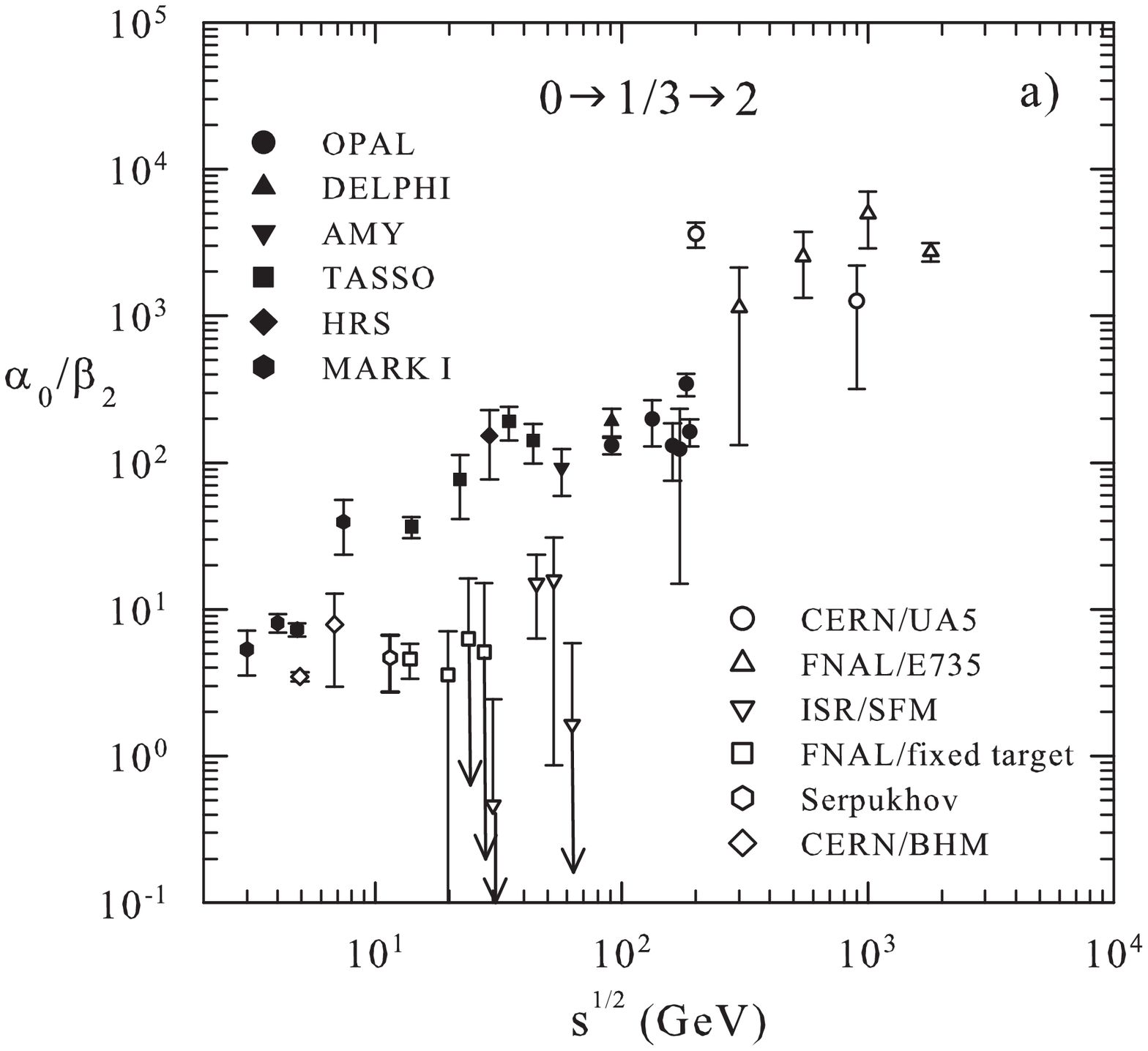}{}}
\hspace*{2.5cm}
\parbox{6cm}{\epsfxsize=5.cm\epsfysize=4.3cm\epsfbox[95 95 400 400]
{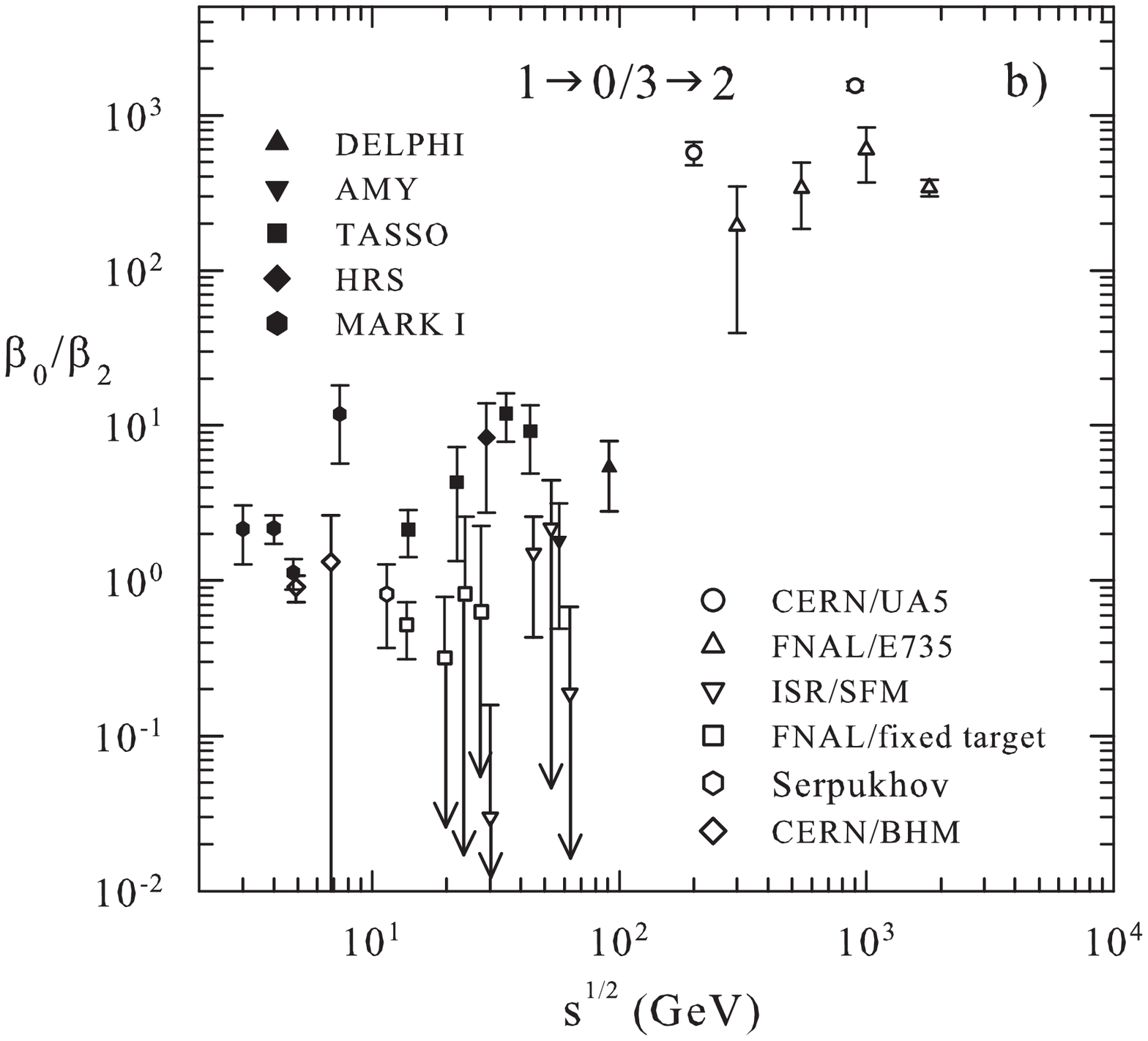}{}}
\vskip 0.6cm
\end{center}
\caption{
The energy dependence of the parameters  $\alpha_0/\beta_2$ and $\beta_0/\beta_2$ 
of GHD obtained from the analysis of data on MD in the full phase-space.
The full and empty symbols correspond to $e^+e^-$ and  $pp/p\bar{p}$ interactions, respectively.  
(a) Ratio of the corresponding rates for the process 
$0\rightarrow 1$ of parton immigration 
and the process  $3\rightarrow 2$  of parton recombination.  
(b) Ratio of the corresponding rates for the process 
$1\rightarrow 0$ of parton absorption and the process $3\rightarrow 2$
of parton recombination. }
\end{figure} 
%
%
\label{Figure:12}
\begin{figure}[] 
\vskip 2.0cm
\begin{center}
\hspace*{-1.5cm}
\parbox{6cm}{\epsfxsize=5.cm\epsfysize=4.3cm\epsfbox[95 95 400 400]
{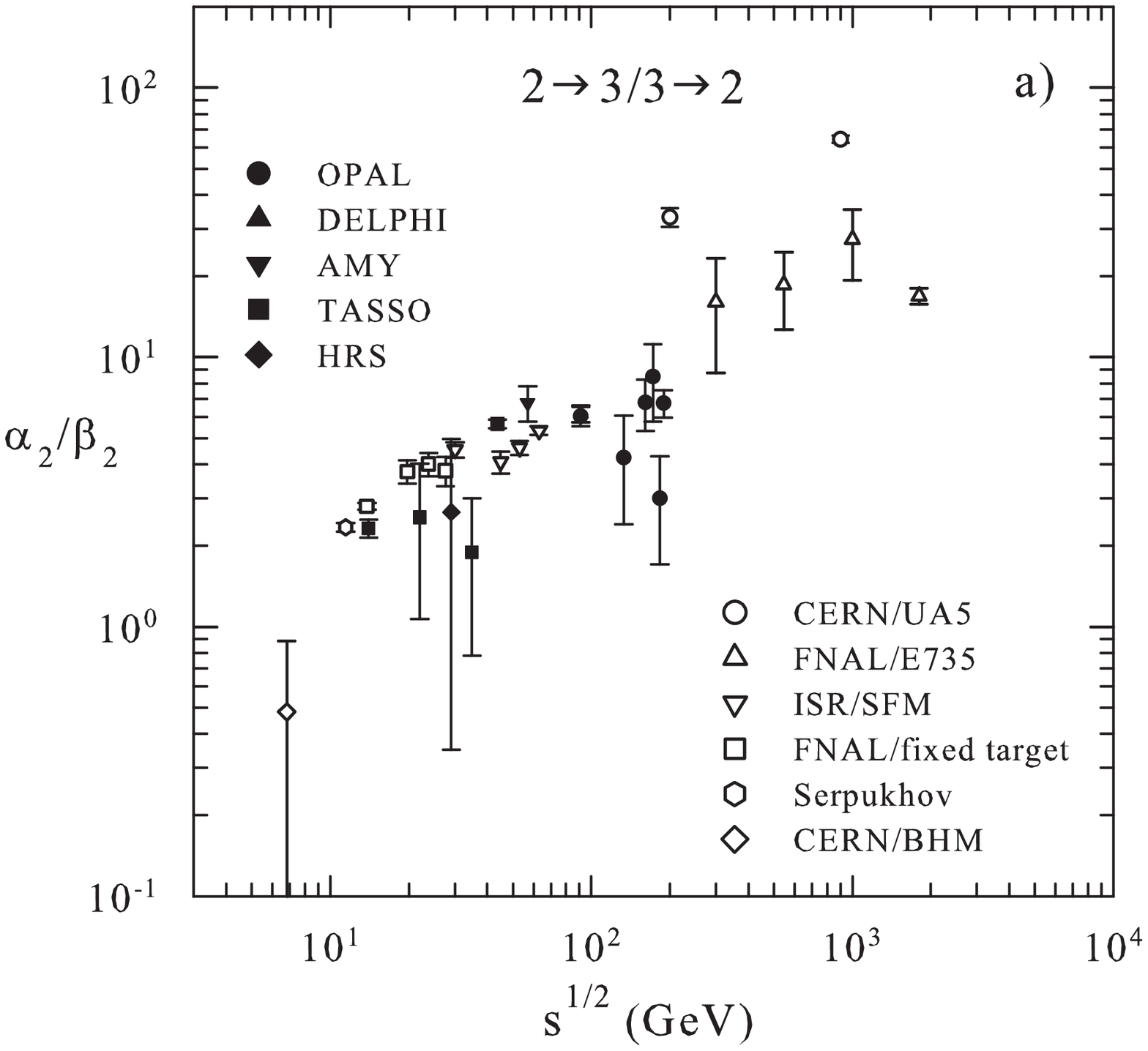}{}}
\hspace*{2.5cm}
\parbox{6cm}{\epsfxsize=5.cm\epsfysize=4.3cm\epsfbox[95 95 400 400]
{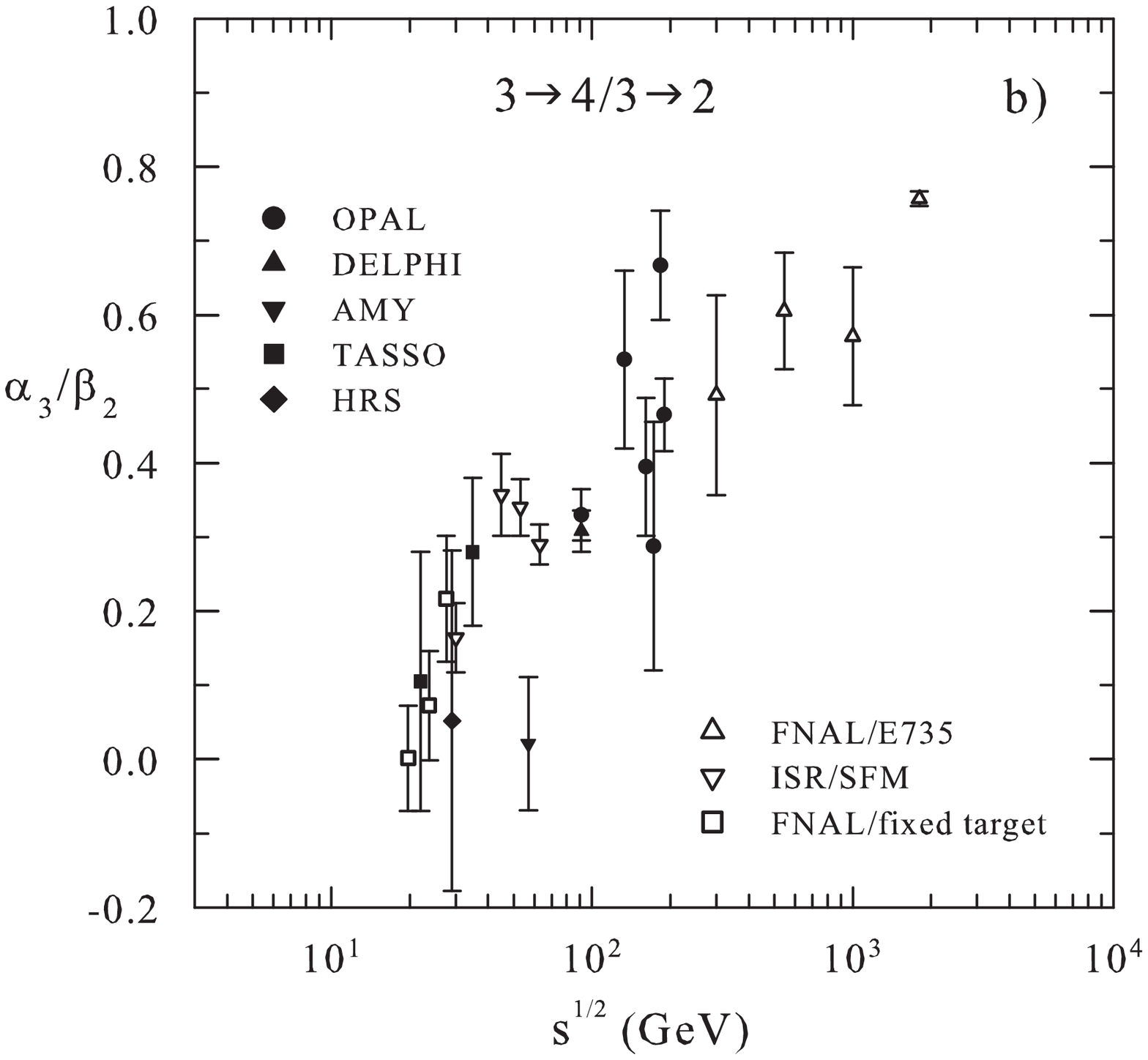}{}}
\end{center}
\caption{
The energy dependence of the parameters $\alpha_2/\beta_2$ and $\alpha_3/\beta_2$ 
of GHD obtained from the analysis of data on MD in the full phase-space.
The full and empty symbols correspond to $e^+e^-$ and  $pp/p\bar{p}$ interactions, respectively.  
(a) Ratio of the recombination rates for the processes $2\rightarrow 3$ 
and $3\rightarrow 2$.  
(b) Ratio  of the recombination rates for the processes $3\rightarrow 4$ 
and $3\rightarrow 2$. } 
\end{figure}

The phenomenological background of data description relies on  
DD evolution equations which include the terms corresponding to the suggested processes.
A stationary solution of the equations gives the recurrence relation (\ref{eq:r10}) 
defining a distribution which we refer to as GHD. 
The essential ingredients of the analysis are four parameters of GHD which are 
the ratios $\alpha_0/\beta_2$, $\alpha_2/\beta_2$, $\alpha_3/\beta_2$, and $\beta_0/\beta_2$  
constructed from the rates $\alpha_i$ and $\beta_i$ of the incremental and decremental 
parton cascade processes, respectively.
The energy dependence of the parameters in the full phase-space which follows 
from the performed analysis 
is shown in Figs. 11 and 12. The empty symbols represent the values obtained from data 
on MD in $pp/p\bar{p}$ interactions. The full symbols correspond to 
$e^+e^-$ annihilations. 
The regular behaviour of all four parameters is guaranteed by the non-zero values of  
$\beta_2$. It means that the recombination process $(3\rightarrow 2)$ is present in both reactions 
at all energies.  
The ratios $\alpha_0/\beta_2$ and $\beta_0/\beta_2$ 
reflect features of data connected with the different initial state 
in the lepton and hadron collisions.
They are proportional to the rates
$\alpha_0$ and $\beta_0$ of the immigration and death processes 
governed by the interactions of partons with the rest of the system as a whole. 
We will refer to them as parameters of the I. type.  
As one can see from Fig. 11(a),  
the parameter $\alpha_0/\beta_2$ increases with $\sqrt{s}$ for both 
reactions. This means that an increase in the collision energy results 
in the relative enhancement of the parton immigration $(0\rightarrow 1)$  
with respect to the three parton decremental recombination $(3\rightarrow 2)$. 
Figure 11(b) shows similar trend in the parameter $\beta_0/\beta_2$ for $pp/p\bar{p}$ collisions.
The relative increase of the rate $\beta_0$ with $\sqrt{s}$ points to  
the intensive absorption of partons in the bulk of the expanding 
system formed in hadron collisions at high energies.   
A different picture is foreseen in $e^{+}e^{-}$ annihilations.  
Here the parameter $\beta_0/\beta_2$ reaches a maximum at $\sqrt{s}\simeq 40$~GeV and    
beyond the energy of $Z^0$ peak it drops to zero (see Table II). 
This suggests that, in contrast to the hadron collisions, the parton absorption  
becomes negligible in the $e^{+}e^{-}$ annihilations at high energies.  

Another difference in the behaviour of the parameters of the I. type 
for the lepton and hadron collisions is in their absolute values. 
In the region below $\sqrt{s}\simeq 40$~GeV, both parameters 
$\alpha_0/\beta_2$ and $\beta_0/\beta_2$ are 
larger for $e^{+}e^{-}$ annihilations in comparison with $pp/p\bar{p}$ interactions.
There are indications from Figs. 11 and 12 that, at high energies, this tendency 
may be opposite. 
A reason for such a behaviour may be influence of jets on the multiplicity structure.
While in the $e^{+}e^{-}$ annihilations at low energies the 
immigration rate $\alpha_0$ stemming from two (or few) jets prevails the parton 
immigration in the hadron collisions, 
the multitude of minijets at high energies would result in much larger $\alpha_0$ in 
the $pp/p\bar{p}$ interactions.     
This in turn would lead to the higher absorption rate $\beta_0$ in the hadron collisions  
giving the partons more probability to be melted conversely in the complex system 
with many minijets again.   

As seen from Figs. 11 and 12, there exists a region in which the parameters 
$\alpha_0/\beta_2$ and $\beta_0/\beta_2$ can be relatively small within the errors indicated.  
The region corresponds to the $pp$ interactions in the energy interval $\sqrt{s}\sim 20-60$~GeV.
The small values of $\alpha_0$ and $\beta_0$ mean that GHD depends mostly on the parameters
$\alpha_2/\beta_2$ and $\alpha_3/\beta_2$. In such case the data can be relatively well approximated by NBD except for a few low values of $n$ (see Fig. 2(b)).   

The energy dependence of $\alpha_2/\beta_2$ and $\alpha_3/\beta_2$
is shown in Fig. 12. 
The ratios depend only on the rates of recombination processes 
which are assumed to be active at the stage of parton-hadron conversions. 
The parameters characterize features of data connected with a breakdown of confinement 
and onset of hadronization. 
We denote $\alpha_2/\beta_2$ and $\alpha_3/\beta_2$ 
as parameters of II. type.
Within the errors indicated, both parameters reveal approximately the same 
energy dependence common for $e^{+}e^{-}$ and $pp/p\bar{p}$ collisions. 
Exceptions make the values for UA5 data \cite{UA5} which result from   
discrepancies at high multiplicities pointed out in Ref. \cite{E735}. 
As seen from Tables I, II and Fig. 12(a), the parameter $\alpha_2/\beta_2$ has a threshold 
in the region $\sqrt{s}\sim 7$~GeV. Afterwards it becomes larger than unity 
and continues in rapid growth at high energies. 
This means that, with increasing $\sqrt{s}$, the rate of the recombination process $(2\rightarrow 3)$ 
prevails still more and more the rate of the inverse process $(3\rightarrow 2)$.     
The parameter $\alpha_3/\beta_2$ has a threshold 
in the region $\sqrt{s}\sim 20$~GeV. 
It grows with the energy and reaches the value of 0.6 at $\sqrt{s}\sim 1$~TeV.
%
%
\label{Figure:13}
\begin{figure}[h!] 
\vskip 2.0cm
\begin{center}
\hspace*{-1.5cm}
\parbox{6cm}{\epsfxsize=5.cm\epsfysize=4.3cm\epsfbox[95 95 400 400]
{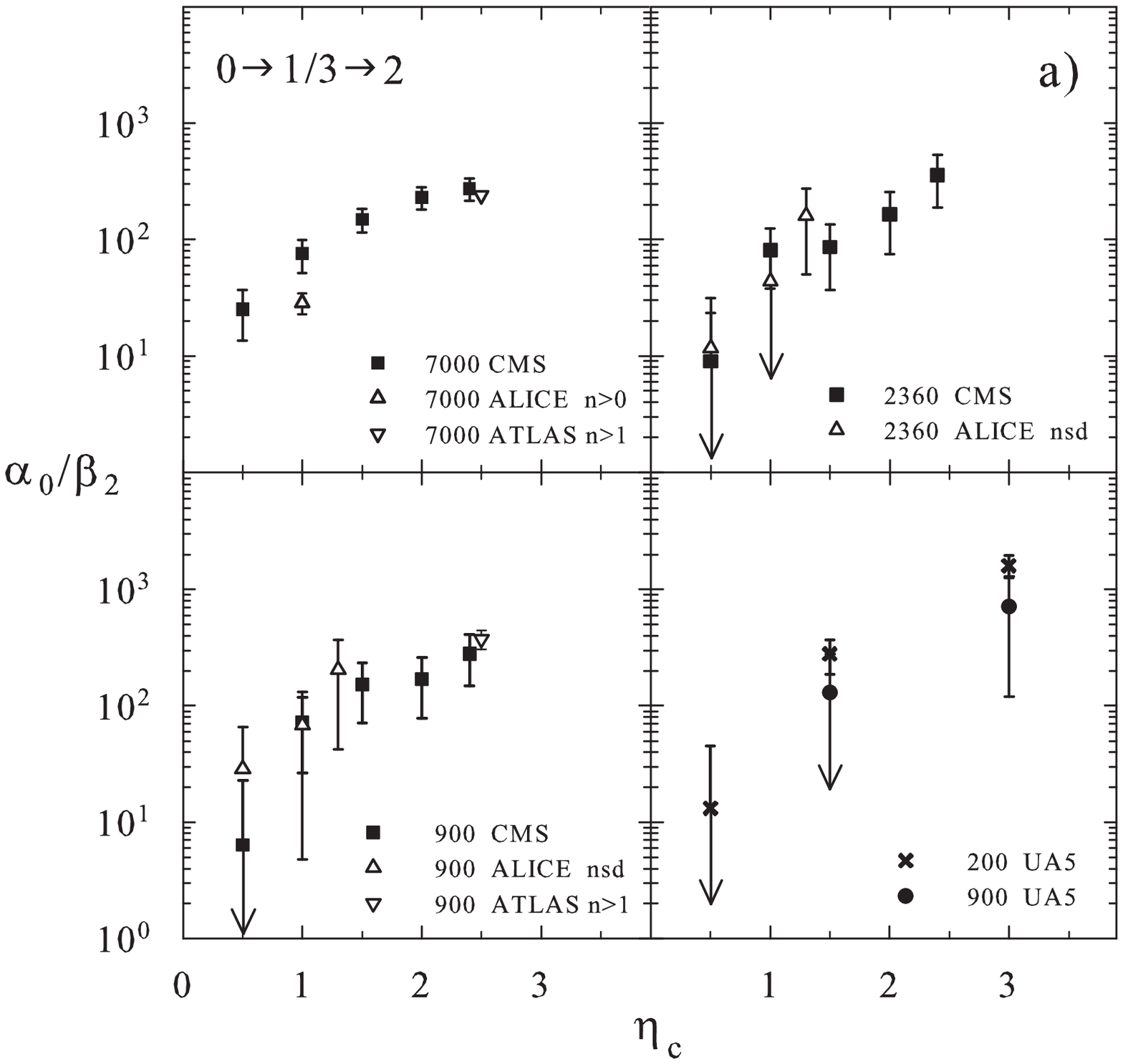}{}}
\hspace*{2.5cm}
\parbox{6cm}{\epsfxsize=5.cm\epsfysize=4.3cm\epsfbox[95 95 400 400]
{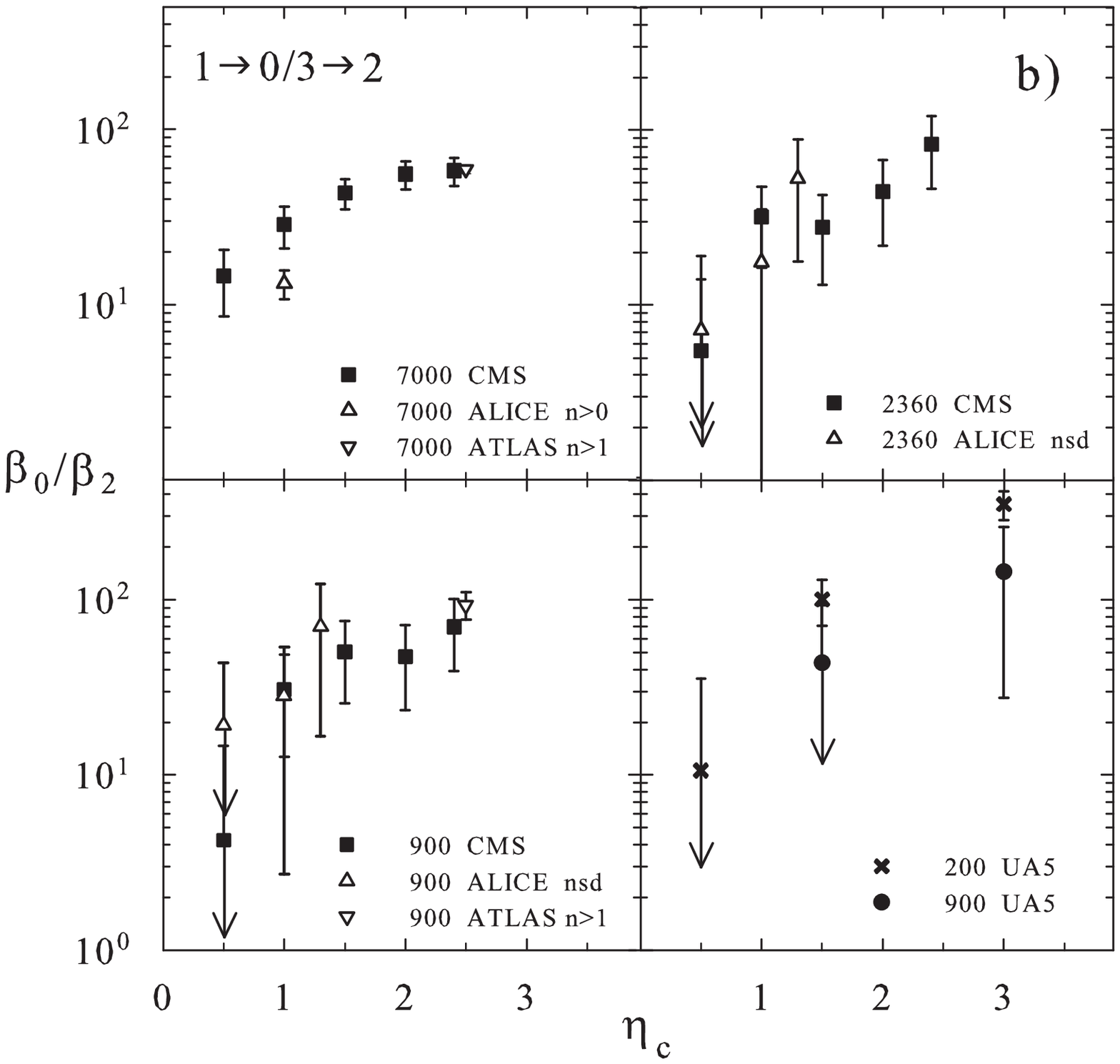}{}}
\vskip 0.6cm
\end{center}
\caption{
The pseudorapidity dependence of the parameters  $\alpha_0/\beta_2$ and $\beta_0/\beta_2$ 
of GHD obtained from analysis of MD in $pp/p\bar{p}$ interactions at different $\sqrt{s}$.
(a) Ratio of the corresponding rates for the process 
$0\rightarrow 1$ of parton immigration 
and the process  $3\rightarrow 2$  of parton recombination.  
(b) Ratio of the corresponding rates for the process 
$1\rightarrow 0$ of parton absorption and the process $3\rightarrow 2$
of parton recombination. }  
\end{figure}
%
%
\label{Figure:14}
\begin{figure}[h!] 
\vskip 1.5cm
\begin{center}
\hspace*{-1.5cm}
\parbox{6cm}{\epsfxsize=5.cm\epsfysize=4.3cm\epsfbox[95 95 400 400]
{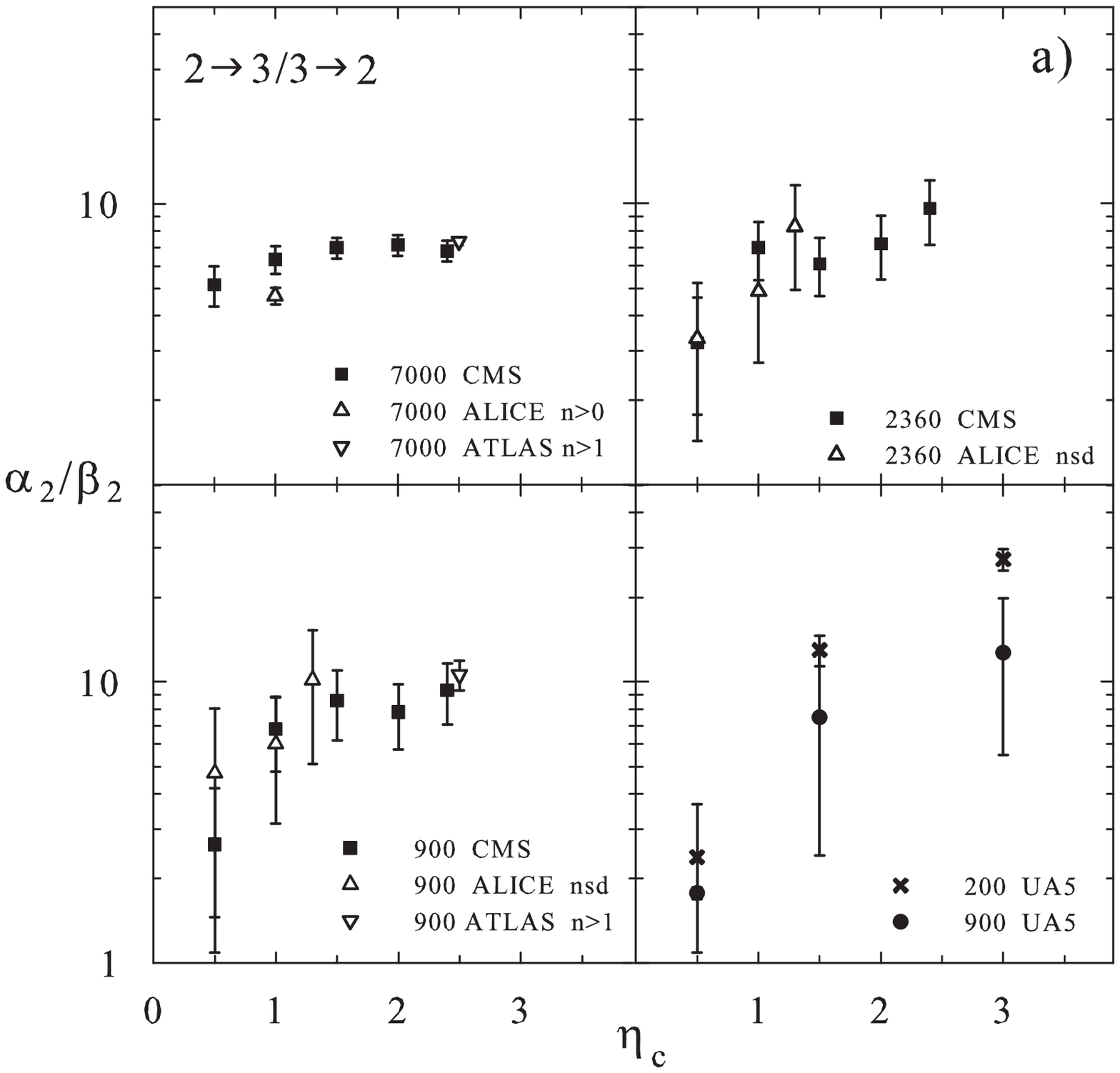}{}}
\hspace*{2.5cm}
\parbox{6cm}{\epsfxsize=5.cm\epsfysize=4.3cm\epsfbox[95 95 400 400]
{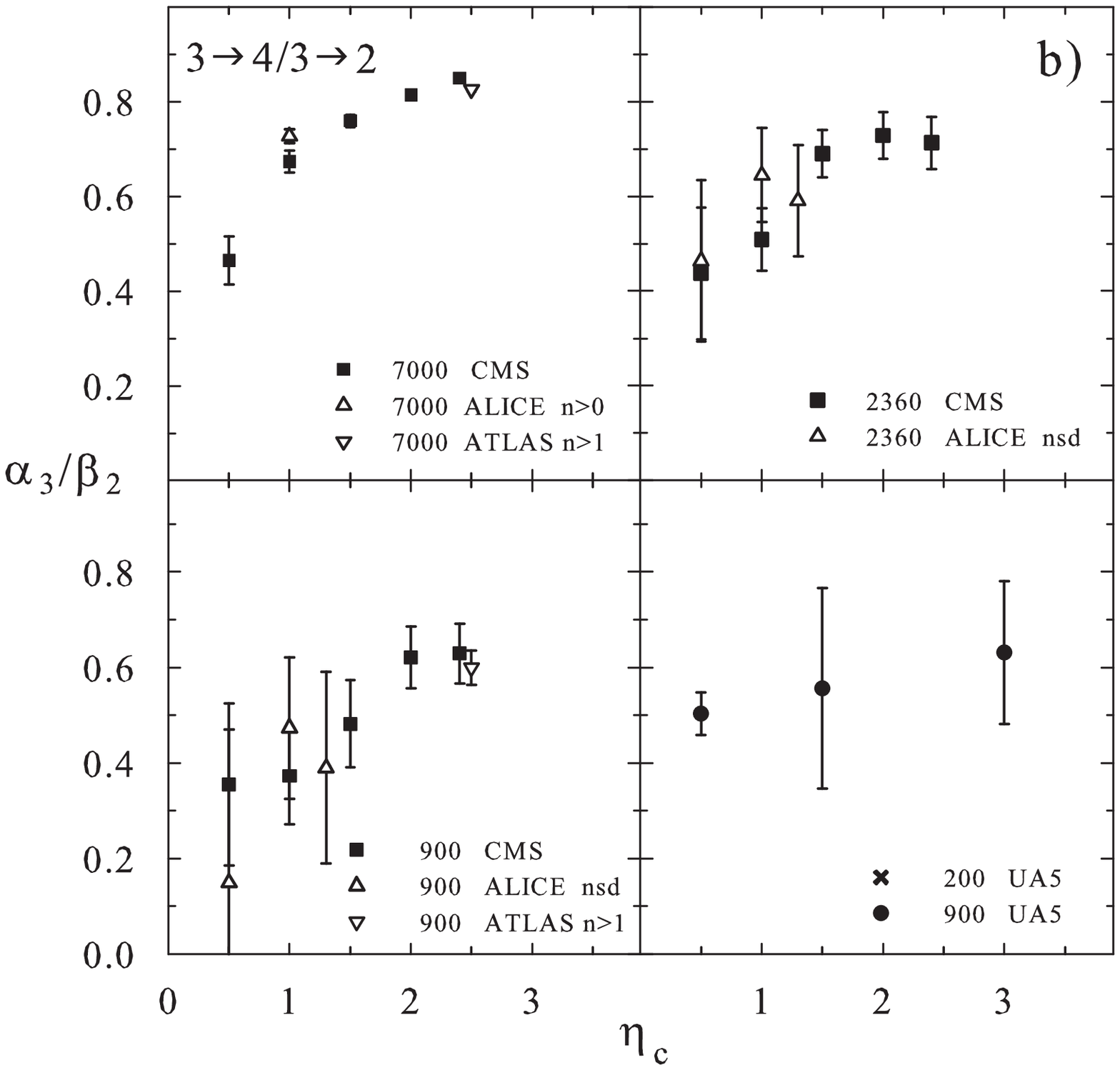}{}}
\end{center}
\caption{
The pseudorapidity dependence of the parameters $\alpha_2/\beta_2$ and $\alpha_3/\beta_2$ 
of GHD obtained from analysis of MD in $pp/p\bar{p}$ interactions at different $\sqrt{s}$.
(a) Ratio of the recombination rates for the processes $2\rightarrow 3$ 
and $3\rightarrow 2$.  
(b) Ratio  of the recombination rates for the processes $3\rightarrow 4$ 
and $3\rightarrow 2$.}
\end{figure}

In contrast to the parameters of I. type, the parameters of II. type 
reflect features of multiplicity production which are common to the 
$e^{+}e^{-}$ and $pp/p\bar{p}$ interactions.      
The analysis suggests that two and three parton recombination processes may be 
part of an intrinsic property of the parton-hadron transitions. 
Physical justification for such an idea may be connected with the 
processes of color neutralization.

Experimental data on MD in the central pseudorapidity windows $|\eta|<\eta_c$ 
allow us to study the behaviour of the parton processes in the limited phase-space regions.
The structure of MD observed in the hadron collisions in the full phase-space demonstrate itself 
distinctly in the limited windows in pseudorapidity 
if the collision energy is sufficiently high.  
This allows more reliable determination 
of the ratios of the corresponding rates of single processes in dependence on the window size.
The fine structure of MD is visible even for small $\eta_c=0.5$ 
in $pp$ collisions at $ \sqrt{s}=7000$~GeV. 
The data at this energy give strongest restriction on the values of the parameters 
in the small pseudorapidity range. 
The dependence of 
the ratios $\alpha_0/\beta_2$, $\alpha_2/\beta_2$, $\alpha_3/\beta_2$, and $\beta_0/\beta_2$  
on $\eta_c$ is depicted in Figs. 13 and 14. 
The symbols represent values of the parameters obtained 
from the analysis of data measured by the CMS, ALICE, ATLAS, and UA5 Collaborations at 
different energies. For clarity reasons, every figure is divided into four panels.
Three panels show the values obtained from analysis of data measured at the LHC at   
$\sqrt{s}=7000$, 2360, and 900~GeV, respectively. The values corresponding to the 
UA5 data are shown on the fourth panel.

One can see from Fig. 13 that $\alpha_0/\beta_2$ and $\beta_0/\beta_2$ increase
with $\eta_c$ at all displayed  energies. Both parameters reveal weak energy dependence 
in the depicted $\eta_c$ region. 
With the increasing window size, the value of $\alpha_0$ becomes still larger than $\beta_0$.
This means that the rate of the immigration process $(0\rightarrow 1)$ grows faster with 
pseudorapidity than the rate of the parton absorption $(1\rightarrow 0)$. 
Except for the energy $\sqrt{s}=7000$~GeV, the parameters 
$\alpha_0/\beta_2$ and $\beta_0/\beta_2$ can be relatively small for $\eta_c=0.5$
within the errors indicated.  
In such a case GHD depends effectively on two parameters of the II. type 
and can be approximated by NBD. 
This is seen in Figs. 4(b), 5(b), and 8(b) where good approximation of data by NBD
at $\sqrt{s}=2360$ and 900~GeV for $\eta_c=0.5$ is demonstrated.   
Such conclusion is in accord with the resume made in Ref. \cite{Mizoguchi}. 
As shown in Fig. 13, the errors of $\alpha_0/\beta_2$ and $\beta_0/\beta_2$ 
at $\sqrt{s}=7000$~GeV exclude small values of these parameters 
even for $\eta_c=0.5$. 
This is why the data cannot be well approximated by NBD.
The statement rephrases the fact that structure of MD stays beyond the NBD 
description in this region (see lowest part of Fig.3(b)).

The pseudorapidity dependence of the parameters $\alpha_2/\beta_2$ and $\alpha_3/\beta_2$
is shown in Fig. 14.  
The ratio $\alpha_2/\beta_2$ reveals growing tendency with $\eta_c$ and flattens at 
$\sqrt{s}=7000$~GeV. This observation suggests importance of the two parton incremental 
recombination process $(2\rightarrow 3)$ in the small pseudorapidity windows 
at this ultra high energy.  
On the contrary, three parton incremental 
recombination $(3\rightarrow 4)$ falls for small $\eta_c$, as shown in the upper left panel
in Fig.14(b). The growing tendency of $\alpha_3/\beta_2$ with $\eta_c$ is seen in the 
LHC data at all
quoted energies.

\section{Conclusions}

Charged particle multiplicity distributions in $pp/p\bar{p}$ collisions have been studied 
including new data from the LHC. The analysis comprises MD in the full phase-space
as well in the limited windows in pseudorapidity. At high energies the distributions 
show a relatively narrow peak
at small multiplicities and a shoulder in the tail.

A phenomenological description of the observed structure was proposed.
Using techniques based on the solution of DD evolution 
equations relevant for the stochastico-physical picture of particle
production a simple formula (\ref{eq:r10}) for the probabilities of
the secondary produced multiplicity has been obtained 
in a stationary regime. 
The basic ingredients of the scenario are the elementary immigration
and absorption of partons and the processes of particle recombination.
We considered two and there particle incremental 
($2\rightarrow 3)$, $(3\rightarrow 4$) and three particle decremental ($3\rightarrow 2$)
recombinations.  
Physical justification for existence of such kind of processes may be connected with  
the requirements of color neutralization at the end of the parton cascade and reaching 
of an approximate "momenta uniformity" of the soft particles at hadronization.  
The features such as two and three parton recombination allow 
to change particle number, exchange particle momenta and
neutralize color repeatedly just before the conversions into the observable hadrons.
The corresponding solution of the higher order equation for the generating function
based on the recombination processes exhibit qualitative properties which
are absent in the first order.

This allowed a quantitative description of the complex structure of data on MD
in $pp/p\bar{p}$ collisions both in the full phase-space and in the limited pseudorapidity 
windows. The phenomenological formula (GHD) was applied to the description of 
the charged particle 
distributions in $e^{+}e^{-}$ annihilations at different energies $\sqrt{s}$. 
A good agreement with data was obtained. 
The dependence of the four parameters of GHD on the energy and pseudorapidity was discussed.
The behaviour of some parameters reveals a universal character  
which is independent of the reaction type while some other parameters depend on it.
It was shown that the incremental recombination processes play an increasingly large role 
in the multiplicity production as the collision energy increases.

Within the approach used and on the basis of the studied
material we conclude that data on MD indicate existence
of certain type of recombination processes correlating particle 
degrees of freedom which manifest itself at high energies.

\section*{Acknowledgments}

The investigations have been supported by the IRP AVOZ10480505, by
the Ministry of Education, Youth and Sports of the Czech Republic
grant LA08002.

\section*{Appendix A}

The system of the evolution rate equations (\ref{eq:r6}) can be solved analytically
in a stationary regime. The stationary solution for the generation function 
satisfies the differential equation (\ref{eq:r8}).  
Introducing the substitution 
\begin{equation}
z=\frac{\alpha_3}{\beta_2}w
\label{eq:a1}
\end{equation}
the equation for $Q(w)=F(z)$ becomes
\begin{equation}
(z^3-z^{2})\frac{d^{3}F(z)}{dz^{3}}  +
\frac{\alpha_2}{\alpha_3}z^{2}\frac{d^{2}F(z)}{dz^{2}} -
\frac{\beta_0}{\beta_2} \frac{dF(z)}{dz} +
\frac{\alpha_0}{\alpha_3}F(z) = 0.
\label{eq:a2}
\end{equation}
It can be rewritten into the known form 
\begin{equation}
\left[\left(z\frac{d}{dz}+a_1\right)
\left(z\frac{d}{dz}+a_2\right)
\left(z\frac{d}{dz}+a_3\right) -
\left(z\frac{d}{dz}+b_1-1\right)
\left(z\frac{d}{dz}+b_2-1\right)\right]F(z) = 0
\label{eq:a3}
\end{equation}
for the generalized hypergeometric function $_3F_2(a_1,a_2,a_3;b_1,b_2;z)$. 
Exploiting the decomposition of  $_3F_2(z)$ into the power series of its argument 
the recurrence relation (\ref{eq:r1}) can be written as follows

\begin{equation}
\frac{(n+1)P_{n+1}}{P_n}=
\frac{(a_1+n)(a_2+n)(a_3+n)}{(b_1+n)(b_2+n)}\frac{\alpha_3}{\beta_2}.
\label{eq:a4}
\end{equation}
Here we used (\ref{eq:a1}) and the standard relation for calculating the probabilities
from the generating function, 
\begin{equation}
P_n = \frac{1}{n!}\frac{d^{n} Q}{dw^{n}}\bigg|_{w=0}.
\label{eq:a5}
\end{equation}
Now, the recurrence relation (\ref{eq:a4}) remains to be expresses in terms of 
the rate parameters $\alpha_i$ and $\beta_i$.   
For this purpose we perform the differentiations in (\ref{eq:a3}) and obtain
 
\begin{eqnarray}
(z^3-z^{2})\frac{d^{3}F(z)}{dz^{3}}  
 + z\left[(3\!+\!a_1\!+\!a_2\!+\!a_3)z-(1\!+\!b_1\!+\!b_2)\right]\frac{d^{2}F(z)}{dz^{2}}
            \nonumber\\
 + \left[(1\!+\!a_1\!+\!a_2\!+\!a_3\!+\!a_1a_2\!+\!a_1a_3\!+\!a_2a_3)z-b_1b_2\right]\frac{dF(z)}{dz}    
           +a_1a_2a_3F(z) =0.            
\label{eq:a6}                     
\end{eqnarray}
The comparison of the coefficients in the equations (\ref{eq:a2}) and (\ref{eq:a6}) gives

\begin{eqnarray}
1+a_1+a_2+a_3+a_1a_2+a_1a_3+a_2a_3 = 0,
            \nonumber\\
   3+a_1+a_2+a_3 = \frac{\alpha_2}{\alpha_3}, \ \ \ \ \
    a_1a_2a_3 = \frac{\alpha_0}{\alpha_3}      
\label{eq:a7}                     
\end{eqnarray}
and
\begin{equation}
1+b_1+b_2 = 0, \ \ \ \ \ b_1b_2 = \frac{\beta_0}{\beta_2}.
\label{eq:a8}
\end{equation}
The first system of equations represents the correspondence between 
the sets of parameters $a_i$ and $\alpha_i$. 
The second system connects $b_i$ and $\beta_i$.
After performing multiplication on the right hand side of (\ref{eq:a4}) and exploiting 
relations (\ref{eq:a7}) and (\ref{eq:a8}), one obtains directly the recurrence relation (\ref{eq:r10}).
  
In the analysis of data we deal with the number of particle pairs $n$. 
The minimal number of the pairs $n_0$ is always greater than null. 
For the NSD $pp/p\bar{p}$ collisions we take $n_0=1$ in all cases. The values of $n_0$ 
for $e^+e^-$ annihilations are shown in Table II. 
The recurrence relation (\ref{eq:r10}) begins always with $n_0$ and the obtained values of 
$P_n$ are renormalized. The corresponding generating function reads 
$Q(w)=cw^{n_0} {_3F_2(z)}$.

\section*{Appendix B}

%
\begin{table}[htb!] 
\caption{The results of analysis of MD of the charged particle pairs in $pp$ 
collisions in the limited phase-space regions $|\eta|<\eta_c$ by the negative 
binomial distribution (NBD) and generalized hypergeometric distribution (GHD). 
The data measured by the CMS Collaboration are taken from Ref. \cite{CMS}.} 
\label{tab:CMS window} 
\begin{center} 
\begin{ruledtabular} 
\begin{tabular}{ccccccccccc} 
$\sqrt{s}$ & $\eta_c$ & NBD & {} & {} & {} &  GHD
\\ 
\cline{3-5}\cline{7-11}
{(GeV)} & {} &  k & q & $\chi^2/\mbox{NDF}$ & {} & 
$\alpha_0/\beta_2$ & $\alpha_2/\beta_2$ & $\alpha_3/\beta_2$ & $\beta_0/\beta_2$ &
$\chi^2/\mbox{NDF}$ 
\\ 
\hline 
7000  & 2.4  & $1.56 \pm 0.03$ & $0.910 \pm 0.001$  & 89.4/67      & {} &
$(2.7 \pm 0.6)10^2$ & $6.8 \pm 0.6$ & $0.85\pm 0.01$ & $58\pm 11$ & 6.4/65
\\ 
  & 2.0   & $1.66 \pm 0.03$ & $0.888 \pm 0.001$  & 140.9/61       & {} &
$(2.3 \pm 0.5)10^2$ & $7.1 \pm 0.6$ & $0.82\pm 0.01$ & $56\pm 10$ & 7.4/59
\\ 
 & 1.5   & $1.72 \pm 0.03$ & $0.851 \pm 0.002$  & 143.5/54       & {} &
$(1.5 \pm 0.4)10^2$ & $7.0 \pm 0.6$ & $0.76\pm 0.01$ & $44\pm 9$ & 3.9/52
\\ 
 & 1.0   & $1.68 \pm 0.04$ & $0.793 \pm 0.003$  & 92.4/38       & {} &
$76 \pm 24$ & $6.4 \pm 0.7$ & $0.67\pm 0.02$ & $29\pm 8$ & 1.9/36
\\ 
  & 0.5   & $1.69 \pm 0.07$ & $0.652 \pm 0.006$  & 51/20       & {} &
$25 \pm 12$ & $5.2 \pm 0.9$ & $0.47\pm 0.05$ & $15\pm 6$ & 1.7/18
\\ 
\hline 
2360  & 2.4  & $1.99 \pm 0.08$ & $0.858 \pm 0.004$  & 30.9/40      & {} &
$(3.6 \pm 1.7)10^2$ & $9.6 \pm 2.5$ & $0.71\pm 0.06$ & $83\pm 37$ & 9.8/38
\\ 
  & 2.0  & $1.89 \pm 0.09$ & $0.840 \pm 0.005$  & 22.2/33      & {} &
$(1.7 \pm 0.9)10^2$ & $7.2 \pm 1.8$ & $0.73\pm 0.06$ & $45\pm 23$ & 8.6/31
\\ 
  & 1.5  & $1.90 \pm 0.09$ & $0.797 \pm 0.005$  & 17.7/27      & {} &
$86 \pm 50$ & $6.1 \pm 1.4$ & $0.69\pm 0.05$ & $28\pm 15$ & 4.4/25
\\ 
  & 1.0  & $2.2 \pm 0.1$ & $0.691 \pm 0.007$  & 33.4/22      & {} &
$81 \pm 43$ & $7.0 \pm 1.6$ & $0.51\pm 0.07$ & $32\pm 16$ &7.2/20
\\ 
  & 0.5  & $1.9 \pm 0.2$ & $0.55 \pm 0.02$  & 5.8/11      & {} &
$9 \pm 14$ & $3.2 \pm 1.4$ & $0.44\pm 0.14$ & $5\pm 9$ &4.6/9
\\ 
\hline 
900  & 2.4  & $2.45 \pm 0.09$ & $0.794 \pm 0.005$  & 28/34      & {} &
$(2.8 \pm 1.3)10^2$ & $9.4 \pm 2.3$ & $0.63\pm 0.06$ & $70\pm 31$ & 6.4/32
\\ 
  & 2.0  & $2.27 \pm 0.09$ & $0.769 \pm 0.006$  & 18.7/31      & {} &
$(1.7 \pm 0.9)10^2$ & $7.8 \pm 2.0$ & $0.62\pm 0.07$ & $47\pm 24$ & 2.9/29
\\ 
  & 1.5  & $2.4 \pm 0.1$ & $0.709 \pm 0.007$  & 23.8/26      & {} &
$(1.5 \pm 0.8)10^2$ & $8.6 \pm 2.4$ & $0.48\pm 0.09$ & $51\pm 25$ & 1.7/24
\\ 
  & 1.0  & $2.6 \pm 0.1$ & $0.603 \pm 0.009$  & 21.6/20      & {} &
$72 \pm 46$ & $6.8 \pm 2.0$ & $0.37\pm 0.10$ & $31\pm 18$ & 2.7/18
\\ 
  & 0.5  & $2.4 \pm 0.3$ & $0.44 \pm 0.02$  & 1.7/10      & {} &
$6 \pm 16$  & $2.6 \pm 1.6$ & $0.36\pm 0.17$ & $4\pm 10$ & 1.1/8
\end{tabular} 
\end{ruledtabular} 
\end{center} 
\end{table}

%
%
\begin{table}[htb!] 
\vspace{-0.5cm}
\caption{The results of analysis of MD of the charged particle pairs in $pp$ 
collisions in the limited phase-space regions $|\eta|<\eta_c$ by the negative 
binomial distribution (NBD) and generalized hypergeometric distribution (GHD). 
The data measured by the ALICE Collaboration are taken from Ref. \cite{ALICE2}.
The NSD data sample is used in the analysis at $\sqrt{s}=2360$ and 900~GeV.} 
\label{tab:ALICE window} 
\begin{center} 
\begin{ruledtabular} 
\begin{tabular}{ccccccccccc} 
$\sqrt{s}$ & $\eta_c$ & NBD & {} & {} & {} &  GHD
\\ 
\cline{3-5}\cline{7-11}
{(GeV)} & {} &  k & q & $\chi^2/\mbox{NDF}$ & {} & 
$\alpha_0/\beta_2$ & $\alpha_2/\beta_2$ & $\alpha_3/\beta_2$ & $\beta_0/\beta_2$ &
$\chi^2/\mbox{NDF}$ 
\\ 
\hline 
7000  & 1.0  & $1.18 \pm 0.02$ & $0.827 \pm 0.002$  & 116.4/37      & {} &
$29 \pm 6$ & $4.7 \pm 0.3$ & $0.73\pm 0.02$ & $13\pm 2$ & 9.8/35
\\ 
\hline 
2360  & 1.3  & $1.75 \pm 0.07$ & $0.782 \pm 0.006$  & 42.7/29      & {} &
$(1.6\pm 1.1)10^2$ & $8.3 \pm 3.4$ & $0.59\pm 0.12$ & $53\pm 35$ & 23.4/27
\\ 
  & 1.0  & $1.67 \pm 0.08$ & $0.742 \pm 0.008$  & 19.6/23      & {} &
$44\pm 45$ & $4.9 \pm 2.2$ & $0.65\pm 0.10$ & $18\pm 18$ & 5.3/21
\\ 
  & 0.5  & $1.66 \pm 0.14$ & $0.576 \pm 0.014$  & 4.9/14      & {} &
$12\pm 20$ & $3.3 \pm 1.9$ & $0.46\pm 0.17$ & $7\pm 12$ & 3.4/12
\\ 
\hline 
900  & 1.3  & $2.15 \pm 0.08$ & $0.702 \pm 0.006$  & 29.9/25      & {} &
$(2.1\pm 1.6)10^2$ & $10.2 \pm 5.1$ & $0.39\pm 0.20$ & $70\pm 54$ & 16.5/23
\\ 
  & 1.0  & $2.00 \pm 0.09$ & $0.656 \pm 0.008$  & 12.3/19      & {} &
$68\pm 64$ & $6.0 \pm 2.9$ & $0.47\pm 0.15$ & $28\pm 26$ & 6.8/17
\\ 
 & 0.5  & $2.14 \pm 0.17$ & $0.458 \pm 0.014$  & 3.9/11      & {} &
$29\pm 37$ & $4.8 \pm 3.3$ & $0.15\pm 0.32$ & $19\pm 24$ & 0.8/9
\end{tabular} 
\end{ruledtabular} 
\end{center} 
\end{table}

%
%
\begin{table}[htb!] 
\vspace{-0.5cm}
\caption{The results of analysis of MD of the charged particle pairs in $pp$ 
collisions in the limited phase-space regions $|\eta|<\eta_c$ by the negative 
binomial distribution (NBD) and generalized hypergeometric distribution (GHD). 
The data measured by the ATLAS Collaboration are from Ref. \cite{ATLAS}.
The sample with the selection $p_T>100$~MeV and $n_{ch}\geq 2$ 
is used in the analysis.} 
\label{tab:ATLAS window} 
\begin{center} 
\begin{ruledtabular} 
\begin{tabular}{ccccccccccc} 
$\sqrt{s}$ & $\eta_c$ & NBD & {} & {} & {} &  GHD
\\ 
\cline{3-5}\cline{7-11}
{(GeV)} & {} &  k & q & $\chi^2/\mbox{NDF}$ & {} & 
$\alpha_0/\beta_2$ & $\alpha_2/\beta_2$ & $\alpha_3/\beta_2$ & $\beta_0/\beta_2$ &
$\chi^2/\mbox{NDF}$ 
\\ 
\hline 
7000  & 2.5  & $1.152 \pm 0.003$ & $0.9263 \pm 0.0003$  & 4397/54      & {} &
$(2.4\pm 0.2)10^2$ & $7.3 \pm 0.3$ & $0.826\pm 0.006$ & $59\pm 4$ & 45.3/52
\\ 
900  & 2.5  & $2.32 \pm 0.02$ & $0.790 \pm 0.002$  & 787/35      & {} &
$(3.7\pm 0.7)10^2$ & $10.6 \pm 1.3$ & $0.60\pm 0.04$ & $94\pm 17$ & 29.2/33
\end{tabular} 
\end{ruledtabular} 
\end{center} 
\end{table}

%
%
\begin{table}[htb!] 
\caption{The results of analysis of MD of the charged particle pairs in $p\bar{p}$ 
collisions in the limited phase-space regions $|\eta|<\eta_c$ by the negative 
binomial distribution (NBD) and generalized hypergeometric distribution (GHD). 
The data measured by the UA5 Collaboration are taken from Ref. \cite{UA5}.} 
\label{tab:UA5 window} 
\begin{center} 
\begin{ruledtabular} 
\begin{tabular}{ccccccccccc} 
$\sqrt{s}$ & $\eta_c$ & NBD & {} & {} & {} &  GHD
\\ 
\cline{3-5}\cline{7-11}
{(GeV)} & {} &  k & q & $\chi^2/\mbox{NDF}$ & {} & 
$\alpha_0/\beta_2$ & $\alpha_2/\beta_2$ & $\alpha_3/\beta_2$ & $\beta_0/\beta_2$ &
$\chi^2/\mbox{NDF}$ 
\\ 
\hline 
900  & 5.0  & $4.26 \pm 0.13$ & $0.799 \pm 0.006$  & 95.2/53      & {} &
$(5.5\pm 4.8)10^3$ & $39 \pm 29$ & $0.3\pm 0.4$ & $(7.7\pm 6.5)10^2$ & 5.8/51
\\ 
  & 3.0  & $2.49 \pm 0.11$ & $0.819 \pm 0.006$  & 25/42      & {} &
$(7.1\pm 5.9)10^2$ & $13 \pm 7$ & $0.6\pm 0.2$ & $(1.4\pm 1.2)10^2$ & 5.6/40
\\ 
  & 1.5  & $2.14 \pm 0.13$ & $0.725 \pm 0.012$  & 13.2/26      & {} &
$(1.3\pm 1.5)10^2$ & $7.5 \pm 5.1$ & $0.56\pm 0.21$ & $(4.4\pm 5.0)10^1$ & 0.6/24
\\ 
  & 0.5  & $1.78 \pm 0.27$ & $0.48 \pm 0.03$  & 0.7/10      & {} &
$(0.2\pm 1.3)10^{-4}$ & $1.77 \pm 0.09$ & $0.50\pm 0.04$ & $(1.3\pm 9.3)10^{-5}$ & 0.3/8
\\ 
\hline 
200  & 5.0  & $5.52 \pm 0.37$ & $0.654 \pm 0.015$  & 7.2/29      & {} &
$(2.9\pm 0.6)10^3$ & $31.5 \pm 2.5$ & {-} & $(4.9\pm 0.9)10^2$ & 0.8/28
\\
  & 3.0  & $4.06 \pm 0.27$ & $0.663 \pm 0.015$  & 8.7/26      & {} &
$(1.6\pm 0.4)10^3$ & $27.3 \pm 2.5$ & {-} & $(3.5\pm 0.7)10^2$ & 0.9/25
\\
  & 1.5  & $3.46 \pm 0.32$ & $0.54 \pm 0.02$  & 3.4/15      & {} &
$(2.8\pm 0.9)10^2$ & $13.0 \pm 1.6$ & {-} & $(1.0\pm 0.3)10^2$ & 1.7/14
\\
  & 0.5  & $5.5 \pm 2.1$ & $0.19 \pm 0.05$  & 0.4/5      & {} &
$13\pm 32$ & $2.4 \pm 1.3$ & {-} & $11\pm 25$ & 0.7/4
\end{tabular} 
\end{ruledtabular} 
\end{center} 
\end{table}

\end{document}